\documentclass[preprint,journal]{vgtc}       
\PassOptionsToPackage{svgnames}{xcolor}
\ieeedoi{10.1109/TVCG.2023.3326574}

\onlineid{0}

\vgtccategory{Research}

\vgtcpapertype{evaluation}

\title{Perceptually Uniform Construction of Illustrative Textures}

\author{%
\authororcid{Anna Sterzik}{0000-0002-0544-5397},
\authororcid{Monique Meuschke}{0000-0003-2183-6619},
\authororcid{Douglas W. Cunningham}{0000-0003-1419-2552},
and \authororcid{Kai Lawonn}{0000-0002-1511-4022} 
}

\authorfooter{
    \item Anna Sterzik and Kai Lawonn are with University of Jena.\\
    E-mail: \{anna.sterzik, kai.lawonn\}@uni-jena.de
    \item Monique Meuschke is with University of Magdeburg. \\
    E-mail: meuschke@isg.cs.uni-magdeburg.de
    \item Douglas W. Cunningham is with Brandenburg University of Technology.
    E-mail: douglas.cunningham@b-tu.de
}

\abstract{%
Illustrative textures, such as stippling or hatching, were predominantly used as an alternative to conventional Phong rendering.
Recently, the potential of encoding information on surfaces or maps using different densities has also been recognized.
This has the significant advantage that additional color can be used as another visual channel and the illustrative textures can then be overlaid.
Effectively, it is thus possible to display multiple information, such as two different scalar fields on surfaces simultaneously.
In previous work, these textures were manually generated and the choice of density was unempirically determined.
Here, we first want to determine and understand the perceptual space of illustrative textures.
We chose a succession of simplices with increasing dimensions as primitives for our textures:
Dots, lines, and triangles.
Thus, we explore the texture types of stippling, hatching, and triangles.
We create a range of textures by sampling the density space uniformly.
Then, we conduct three perceptual studies in which the participants performed pairwise comparisons for each texture type.
We use multidimensional scaling (MDS) to analyze the perceptual spaces per category.
The perception of stippling and triangles seems relatively similar.
Both are adequately described by a 1D manifold in 2D space.
The perceptual space of hatching consists of two main clusters:
Crosshatched textures, and textures with only one hatching direction.
However, the perception of hatching textures with only one hatching direction is similar to the perception of stippling and triangles.
Based on our findings, we construct perceptually uniform illustrative textures.
Afterwards, we provide concrete application examples for the constructed textures.

\vspace{1em} 
  \noindent \textbf{Correction Note (Dec 2025):} 
In this author version, we have corrected an error present in the official IEEE publication. Specifically, the x and y axis labels in Figures 12 and 13 were swapped, which previously led to incorrect instructions regarding the application of the sigmoid function versus its inverse.
}


\teaser{
    \centering
    \includegraphics[width = \linewidth]{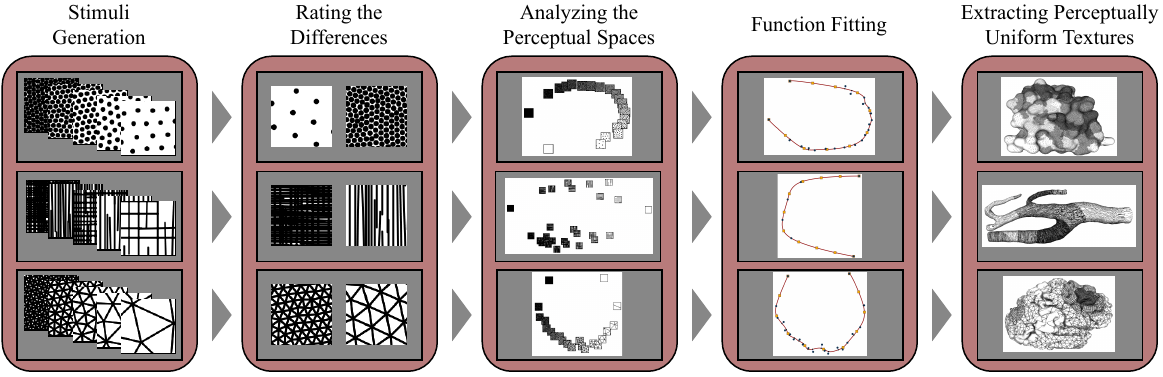}
    
    \caption{Workflow of our evaluations. First, the stimuli were created.
    Then, in three crowdsourced, online studies, participants rated the differences between textures of different densities.
    The results of the studies were analyzed using multidimensional scaling (MDS).
    We constructed 2D embeddings for the textures---the perceptual spaces.
    Afterwards, we reparameterized the perceptual spaces using function fitting.
    Finally, we extracted perceptually uniform textures.}
    
    \label{fig:workflow}
}

\graphicspath{{figs/}{figures/}{gfx/}{pictures/}{images/}{./}} 

\usepackage{todonotes}
\usepackage{svg}
\usepackage{booktabs} 
\usepackage{float}
\usepackage{comment}
\usepackage{mathptmx}                  
\usepackage{amsmath}
\usepackage{mathtools}
\usepackage{placeins}
\newcommand{\review}[1]{#1}
\begin{document}



\firstsection{Introduction}
\maketitle

The encoding of data on surfaces is an important topic in visualization.
By visually encoding complex data, it can be displayed in an intuitive manner and underlying patterns and relationships can be revealed.
Often, the encoded data are scalar fields.
Over several decades, many techniques and algorithms were developed to display scalar data on surfaces.

Color mapping is one of the most popular and widely used of these techniques.
However, color mapping has some shortcomings.
Neither is it accessible to all viewers nor does it allow the simultaneous display of multiple scalar fields at once.
Recently, illustrative textures---predominantly stippling and hatching---have been proposed for information encoding~\cite{Gortler2019, Taylor2002, Lawonn2017, Lawonn_2015_MICCAI, Meuschke_2017_TVCG}.
Traditionally, illustrative visualization techniques were rather used as an alternative to Phong shading.
Higher densities of the primitives---dots and dashes for stippling and hatching respectively---correspond to areas with darker shading on the model.
For information encoding, higher densities can be used to encode higher data values.
While stippling can only represent one scalar field at a time, crosshatching can be used to display two types of data simultaneously.
For example, Sajadi et al.~\cite{Sajadi2013} used positive slopes to indicate a share of green and negative slopes for a share of red to help deuteranopes with retrieving color information.

In previous work, the density levels were determined unempirically~\cite{Lawonn2017, Lawonn_2015_MICCAI, Meuschke_2017_TVCG}.
\review{
However, it is unclear if the perception of changes in the textures is similar to the physical changes.
}
For the more complex crosshatching, an additional question arises:
Previously a succession of textures that begins with horizontal lines and later adds vertical hatches has been used for scalar information encoding \cite{Meuschke_2017_TVCG}.
This is a classic approach in artistic drawings but is it intuitive for information encoding?
In this paper, we explore the perceptual spaces of illustrative textures to facilitate an informed selection of density levels.

As a structured approach to selecting texture types for the evaluation, we chose to investigate simplices with increasing dimensions as primitives.
Dots for stippling, lines for hatching, and triangles for triangulated textures.
For each texture type, we sampled the density space linearly.
From no primitives to completely covered by primitives.
To analyze the perceptual spaces of the so-created textures, we conducted three crowdsourced, perceptual studies with 60 participants.
In these studies, the participants were asked to rate the pairwise differences between all textures of the same type.
We analyzed the resulting similarity matrices by performing multidimensional scaling (MDS).
For a given N $\times$ N matrix of similarity ratings for N objects and a dimension d, MDS approximates the coordinates in d-dimensional space.

We found that the perception of stippled and triangulated textures is adequately described by a 1D manifold in 2D space.
The perception of crosshatched textures is more complex.
While there exists a reasonable 2D representation, it is not possible to adequately describe the perceptual space of hatching in 1D or with a 1D manifold.
One of the most distinctive features of the perceptual space of hatching is a clustering of the textures into two main groups: crosshatched textures and textures with only one hatching direction.
The perceptual distances between those two clusters are rather large.
Thus, we would advise against a mixed use of one-directional hatching and crosshatching for encoding a scalar field.
The perception of one-directional hatching seems to be similar to the perception of stippling and triangulated textures.
We reparameterized the three perceptual spaces to construct perceptually uniform textures.
We found that the perception of illustrative textures is highly non-linear around the extreme ends of the density spectrum.
%
Our reparameterizations suggest that sigmoid functions provide an adequate mapping from the perceptual space to the original density values.
%
Finally, we present application examples for perceptually uniform illustrative textures.

To summarize, we provide a structured evaluation of illustrative textures for information encoding.
We conducted three crowd-sourced studies to explore the perceptual spaces of stippling, hatching, and triangulated textures.
By analyzing the perceptual spaces, we can give guidance on constructing perceptually uniform density levels.

\section{Related Work}
In the following Section, we describe related work in illustrative visualization (Section \ref{sec: rw IlluVis}), approaches for encoding scalar data on surfaces (Section \ref{sec: rw data encoding}), and perceptual research (Section \ref{sec: rw perceptual research}).
\subsection{Illustrative Visualization}
\label{sec: rw IlluVis}
Illustrative visualization techniques are an alternative to traditional rendering.
The goal is to convey information without the need for super-realistic representations.
In addition to providing an alternative to traditional rendering, illustrative visualization can also be used for focus and context rendering~\cite{Lawonn_2014_BVM,Lawonn2014}, and as an additional visual channel for encoding information on the surface~\cite{Meuschke_2017_TVCG}. 
There is a plethora of different illustrative visualization techniques that can be categorized into silhouettes and contours, feature lines, stippling, hatching, and shading~\cite{Lawonn2018}.
In the following, we will focus on the categories of stippling and hatching as these cover the focus of this paper.
For more information on silhouettes, contours, and feature lines, we refer to the following surveys~\cite{Isenberg2003,Lawonn_2015_Feature}.

These techniques can be further divided into image-based, texture-based, and object-based techniques.
Image-based techniques work with the resulting image, i.e. a domain of pixels and colors.
Contour lines may for example be generated based on a shaded image.
Texture-based techniques use the surface mesh and parameterize it to obtain the texture coordinates needed for texturing.
Finally, object-space techniques also use surface meshes but generate illustrative visualization techniques as geometric primitives. 
For our work, the texture-based techniques are most relevant as we use textures to explore perceptual spaces.

The hatching technique is about placing lines on a surface to convey its shape.
Stippling techniques try to do the same but with points instead of lines.
One of the first hatching approaches was presented by Interrante et al.~\cite{Interrante1996}, who discussed using the principal curvature direction as the direction for the hatching strokes.
Hertzmann and Zorin~\cite{HertzmannZorin2000} presented another seminal hatching approach.
They generated streamlines as hatching strokes but also ensured that the underlying vector field was smoothed, yielding visually pleasing results.
Besides generating single lines, Praun et al.~\cite{Praun2001} suggested using texture patches with mapped hatch strokes.
These textures covered the surface and provided an illustration of a mesh with a hatching style.
We utilize their method of distributing the line primitives in a visually pleasant way (see Section \ref{sec: pe stimuli}).
Other approaches that compute line primitives directly on the surface were provided by Lawonn et al.~\cite{Lawonn2013} and Gerl and Isenberg~\cite{Gerl2013}.
A limitation of the first approach was that the techniques were tesselation dependent and the hatching could not be applied to animated surface meshes in real-time.
This limitation has been addressed by a novel hatching technique presented by Lawonn et al.~\cite{Lawonn2014}.

Unlike hatching, stippling uses dots instead of lines.
Secord et al.~\cite{Secord2002} presented an approach to distribute primitives by providing an image.
A probability density function based on tone is used to determine how many primitives should be generated in which regions. 
Krüger and Westermann~\cite{Kruger2007} developed a stippling approach by generating a 3D noise volume that is superimposed on the surface to assign points based on the underlying shading.
Based on properties of an underlying image, e.g., grayscale, variance, Deussen et al.~\cite{Deussen2017} placed dots using an extension of Lloyd's optimization method.
This is similar to our approach for generating stippling and triangles.
For an overview of stippling techniques, see the review by Martìn et al.~\cite{Martin2017}.

Stoppel et al.~\cite{Stoppel2019} developed an algorithm that uses the tonal content of an image to create illustrative patterns that mimic the image.
In their work, they present different styles that give visually pleasing results.
Besides others, they feature the techniques that we used in our studies: stippling, hatching, and triangulation.
Lawonn and Günther presented an algorithm on the GPU to cover a given image with triangles~\cite{Lawonn2019_Tri}.
This approach was motivated by image stylization. 
Later, Xiao et al.~\cite{Xiao2022} extended the image triangulations by adding curved edges on the triangles to ensure that the triangles are better fitted to the image.

\subsection{Encoding Information on Surfaces}
\label{sec: rw data encoding}
Visualizing scalar fields on surfaces has been a popular topic in visualization for decades, with various techniques proposed in domains such as medical imaging, geology, meteorology, and fluid dynamics.

One of the most widely used techniques is the isocontour extraction method, which involves extracting and visualizing isocontours, or curves of constant scalar value, on the surface~\cite{liu2010construction}. Several algorithms have been developed to efficiently extract isocontours from volumetric datasets, such as the marching cubes algorithm and its variants.
Isocontours are an alternative to our texture-based encoding. They can also be used in addition to color-coded surfaces.

Another typical technique is to color-encode scalar values on surfaces~\cite{Coninx2011}. Here, either discretized or continuous color scales can be employed depending on the analysis task. While the rainbow color scale is still chosen in many applications, several problems occur with this scale such as artifacts or less contrast for high-frequent data. Therefore, perceptual-oriented color scales such as the Viridis or temperature scale should be used.
We propose to use texture-based methods as either an enhancement or an alternative to color-coded data.
Furthermore, textures-based methods could be used as an additional channel to encode further data.
Coninx et al.~\cite{Coninx2011} used Perlin noise textures to encode uncertainty. Perlin noise is a type of gradient noise to create natural-looking textures and patterns. It is generated by combining multiple layers of smoothly interpolated random values, or gradients, at different scales and orientations. The resulting noise function has a smooth and continuous appearance, with no abrupt changes or repeating patterns. The smoothness and randomness of the noise function can be adjusted by controlling the number and size of the gradient layers, as well as the interpolation function used to combine them. 

Furthermore, illustrative techniques can be used to visualize scalar fields on surfaces. 
Lawonn et al. used different dense stippling~\cite{Lawonn_2015_MICCAI} and hatching patterns~\cite{Lawonn2017} to encode the distance between a tumor and the surrounding vasculature. Lichtenberg et al. developed a surface parametrization for tree-like structures, which is used to generate varying contour renderings to encode the distance to a tumor. Sterzik et al.~\cite{Sterzik2022, Sterzik2023} used line renderings with varying line attributes to encode uncertainties on molecular surfaces.

Besides the encoding of a single scalar field, multiple scalar fields can be encoded on surfaces by combining methods to visualize scalar information. Meuschke et al.~\cite{Meuschke_2017_TVCG} used a combination of color and image-based hatching to encode
two scalar fields on 3D surface meshes representing cerebral aneurysms. Lawonn et al.~\cite{Lawonn_2014_CGFb} employed a 2D linear color scale to visually 
highlight concave and convex regions on vascular surfaces.
Taylor~\cite{Taylor2002} used sets of dots and bumps or differently oriented lines for representing several scalar fields.
Different colors and scales or intensities were used to encode the scalar value.

\subsection{Perceptual Research}
\label{sec: rw perceptual research}
In the last decades, many low-level psychology research works investigated the visual perception of textures. 
Lettvin et al.~\cite{lettvin1976} proposed that texture is the fundamental element from which form perception is constructed, 
taking into account numerous detailed observations on peripheral form vision. 
Julesz et al.~\cite{caelli1978perceptual1,julesz1981,julesz1973inability} further developed this proposal by demonstrating that texture perception disregards relative spatial position, unlike form perception, which is based on local scrutiny. 
In contrast, Cunningham et al.~\cite{cunningham2007perceptual} performed more high-level psychology research.
They investigated the difficulty in determining the relationship between the parameters of a computer graphics algorithm and the resulting perceptual effect, which is important for scientific and artistic endeavors. 
The authors propose a generalized method using MDS and Factor Analysis to determine the qualitative and quantitative mapping between parameters and perception. 
The method is demonstrated using two datasets of glossy and transparent objects, revealing the aesthetic and low-level material properties represented by the factors. 
The authors suggest that perceptual reparameterization of computer graphics algorithms can improve their accessibility, allowing for easier generation of specific effects and more intuitive exploration of different image properties.
Wills et al.~\cite{Wills2009} investigated the perceptual space of gloss.
They constructed a perceptually uniform sampling by interpolating between several bidirectional reflectance distribution functions in the perceptual space.
Similarly, our aim is the reparameterization of the perceptual space of different illustrative textures, which is why we do not focus on low-level psychology research. 

Lie et al.~\cite{liu2015visual} discussed the limitations of using procedural models to generate textures in computer graphics, which lack intuitive and perceptual characteristics. 
They conducted two psychophysical experiments to investigate the perceptual dimensions of textures and used hierarchical cluster analysis and singular value decomposition to identify the features used by observers in grouping similar textures. 
They found that existing dimensions in literature cannot accommodate random textures, and used isometric feature mapping (Isomap) to establish a three-dimensional perceptual texture space. 
Finally, they proposed computational models to map perceptual features to the perceptual texture space, which can suggest a procedural model to produce textures based on user-defined perceptual scales.

Spicker et al.~\cite{Spicker2017} investigated the perception of the abstraction quality of stippled images based on the number of stipples.
Martìn et al.~\cite{Martin2019} analyzed the drawing characteristics of hand-drawn stipples.
Furthermore, they evaluated the perception of digital stipples to extract guidelines for reproducing hand-drawn stipples. 
Görtler et al.~\cite{Gortler2019} used stippling to encode data on maps.
They conducted a perceptual study to determine the just-noticeable differences (JND) for grayscale and three types of stippled textures.
Sterzik et al.~\cite{SterzikLines} used JND evaluations to analyze the perception of line attributes for encoding data in visualizations.
They first identified popular line attributes by letting people draw line-based visualizations.
Afterwards, they investigated discriminability and gave level recommendations for the line attributes based on evaluating the JND.
JND can also serve as a basis for sampling perceptually uniform textures.
\review{We chose an MDS-based approach instead of a JND-based approach because MDS analysis allows us to efficiently gain a more complete overview of the metric structure of the whole perceptual space.
The reconstruction of the whole space is---by definition global---which is not the case with the purely local approach of JNDs.
}
In our evaluation, we have included stippling to provide a more comprehensive overview of illustrative textures, as using the same study setup allows for better comparability of results.
Moreover, we can examine whether our MDS evaluation replicates the conclusions drawn by Görtler et al. regarding stippling perception, such as its non-linear nature and the need for root or logarithmic scaling of density mapping.
To the best of our knowledge, no existing research has calculated JND for other illustrative textures.
The results of two user studies by Sajadi et. al.~\cite{Sajadi2013} show, that the use of different patterns, including but not limited to hatching, in addition to color, could help dichromats with colormap interpretation and categorization.

\section{Perceptual Evaluations}
We attempt to analyze the perceptual spaces of stippled, hatched, and triangulated textures.
Figure \ref{fig:workflow} shows the workflow for the generation of perceptually uniform textures.
First, we generated illustrative textures as stimuli for the evaluations (Section \ref{sec: pe stimuli}).
Afterwards, we conducted a crowdsourced, perceptual study for each texture type.
In these studies, the participants rated the pairwise differences for each pair of textures per texture type.
The study procedure is described in Section~\ref{sec: pe method}.
Then, we analyzed the study results with multidimensional scaling (Section \ref{sec: pe analysis}.
In Section \ref{sec: construction}, we use the results of our perceptual evaluations to construct perceptually uniform texture levels.
Finally, in Section \ref{sec: application}, we show application examples for our texture levels.
\begin{figure}[tb]
    \centering
    \scriptsize
     \includegraphics[width = \linewidth]{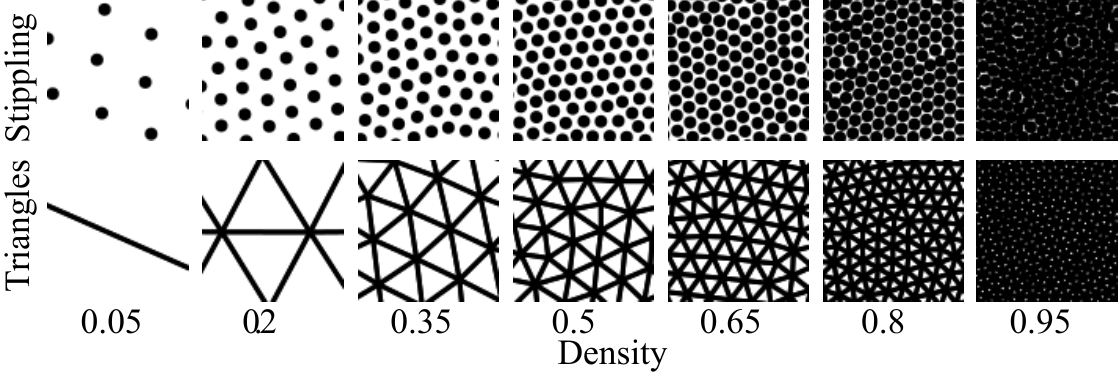}
    \caption{Subset of the stippling (first row) and triangle (second row) stimuli used in the perceptual studies. The images have been cropped and scaled down. The whole stimuli range, ranges from empty (0) to completely covered (1) and is sampled linearly with a step size of 0.05.}
    \label{fig:stimuli_stipple_triangle}
\end{figure}
\begin{figure}[tb]
    \centering
    \scriptsize
    \includegraphics[width=0.75\linewidth]{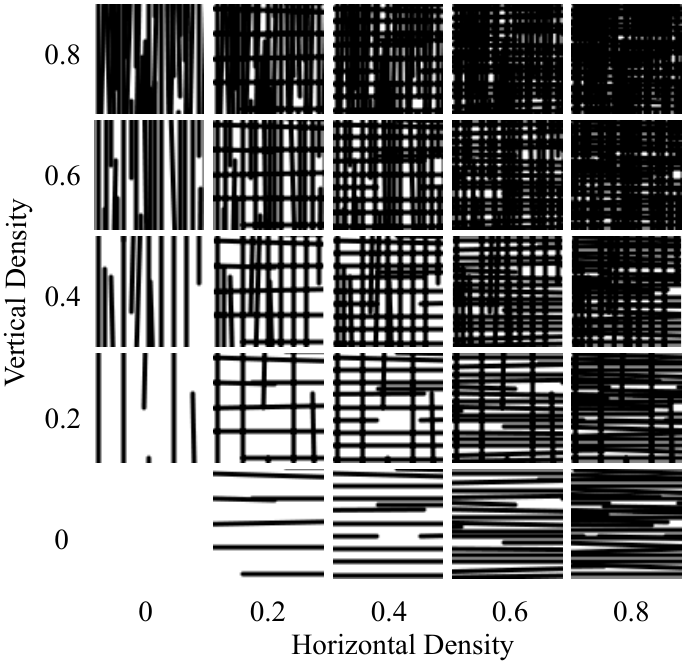}
    \caption{Hatching stimuli used in the perceptual studies. The images have been cropped to fit the page. In the experiment, an additional black texture (full density, completely covered) was present.}
    \label{fig:stimuli_hatching}
\end{figure}

\subsection{Stimuli Generation}
\label{sec: pe stimuli}

The stimuli for each texture type were constructed similarly.
First, we sampled the density space.
The density values range from 0 (empty) to 1 (completely covered by primitives).
For the density values in-between, we count the pixels covered by primitives and compute the mean.
For black primitives on a white background, this is similar to $1 - \textrm{mean gray value}$.
For the generation of stippled and triangulated textures we only needed to vary one parameter, therefore, we chose stepsizes of 0.05.
For crosshatching we needed to consider horizontal and vertical hatches, therefore the sampling needed to be a lot coarser.
To make the experiments feasible we chose a stepsize of 0.2
The generation of stimuli textures was an iterative process---we added more primitives until the desired density was reached.
For hatching, the horizontal and vertical textures were generated individually.
In the second step, we created overlays for every pair of vertical and horizontal hatches.
Each texture had a size of $512 \times 512$ pixels.\\

\noindent\textbf{Stippling.}
For stippling, in each iteration, we positioned a new stipple dot randomly in the texture. 
Each stipple dot had a diameter of 8 pixels.
This corresponds to the stipple size for which Görtler et. al.~\cite{Gortler2019} found the most linear perception.
Similar to Deussen et al.~\cite{Deussen2017}, we distributed the stippling points equally in the image by using Lloyd's algorithm~\cite{Lloyd1982}.
The first row of Figure \ref{fig:stimuli_stipple_triangle} shows a subset of the stippling textures used during the study.\\

\noindent\textbf{Triangulated Textures.}
The generation of triangulated textures is similar to the generation of stippled textures.
New points are added iteratively and are then distributed with Lloyd's algorithm.
In each iteration, a Delaunay triangulation is generated for the distributed points.
The triangles are drawn with a line width of 3 pixels.
This line width strikes a balance between thickness and thinness, enabling the generation of finely detailed density levels without sacrificing clarity. 
The second row of Figure \ref{fig:stimuli_stipple_triangle} shows a subset of the triangle textures used during the study.\\

\noindent\textbf{Hatching.}
For the hatching stimuli, the parameter space is two-dimensional.
\review{
For each combined texture, we sampled two individual textures, one with only horizontal and one with only vertical lines.
The stepsizes between the sampled densities were 0.2.
We create the final 2D space by overlaying horizontal and vertical lines (Figure \ref{fig:stimuli_hatching}).
Each crosshatched texture is defined by two density values.
The first value describes the density of the texture with only horizontal dashes, and the second value is the density of the texture with only vertical hatches.
The density of the resulting crosshatched texture cannot be obtained by simply averaging the densities in each dimension.
If this density needs to be calculated, the number of pixels covered by the primitives must be determined for the complete crosshatched texture without considering individual dimensions.
}
We used the method of Praun et al.~\cite{Praun2001} to generate equally spaced hatching lines.
In each iteration, several candidate lines were generated.
Again, each line is 3 pixels wide and has a randomly chosen length between 0.3 times the image width and the full image width.
For each resulting candidate texture, an image pyramid was created.
Then, the sum of density added by the candidate stroke over all images in the pyramid was computed.
The results were then divided by the line length for a measure of the goodness of the line.
The line with the highest goodness was chosen from the number of candidates.
This process prevents the lines from forming clusters. \\

\begin{figure}[tb]
    \centering
    \includegraphics[width=0.75\linewidth]{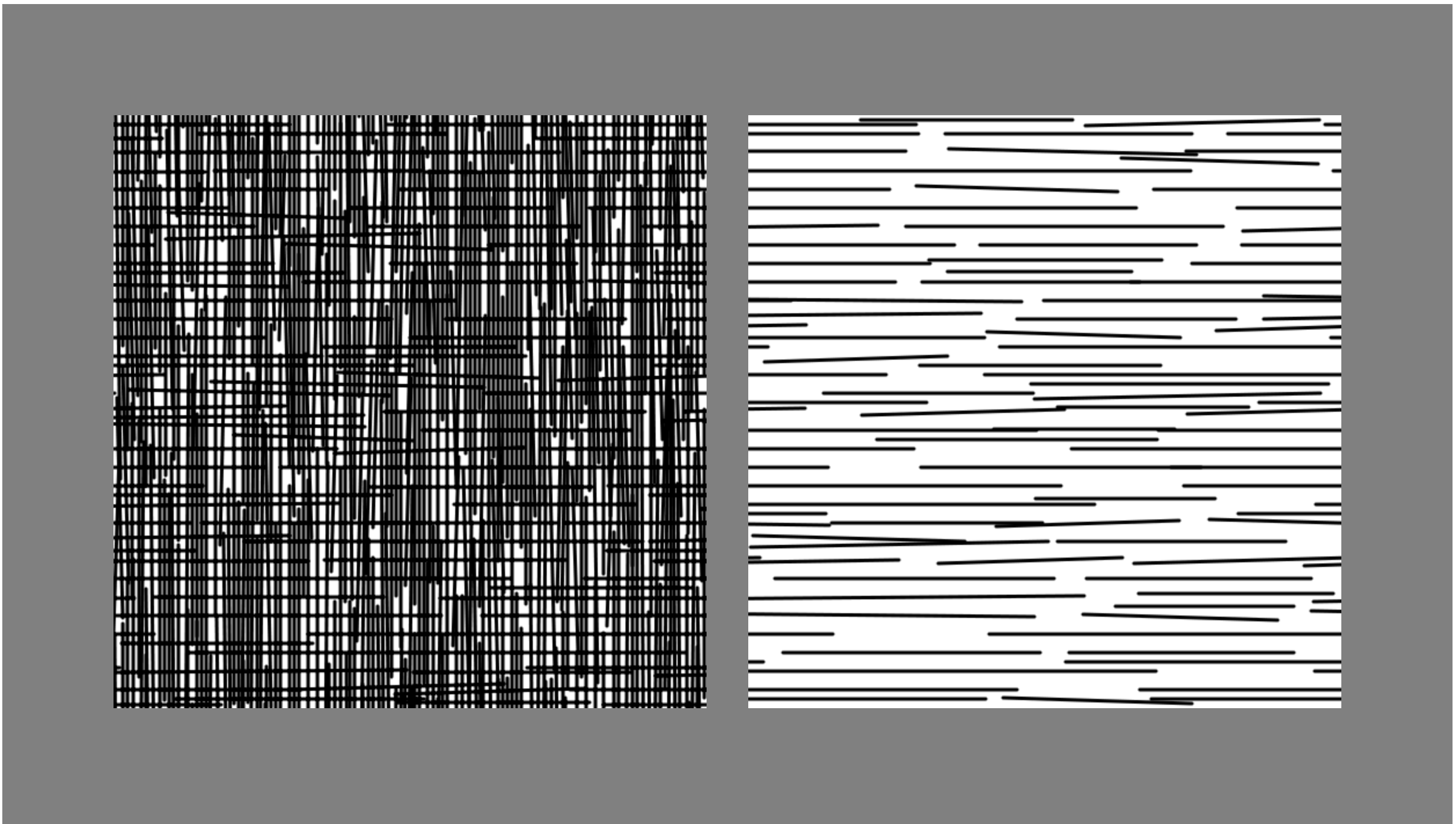}
    \caption{Example task of the hatching study.
    The participants rated the difference from 1 (very similar) to 9 (very different).}
    \label{fig:example_screen}
\end{figure}
\begin{figure}[tb]
    \centering
    \scriptsize
    \includegraphics[width=0.84\linewidth]{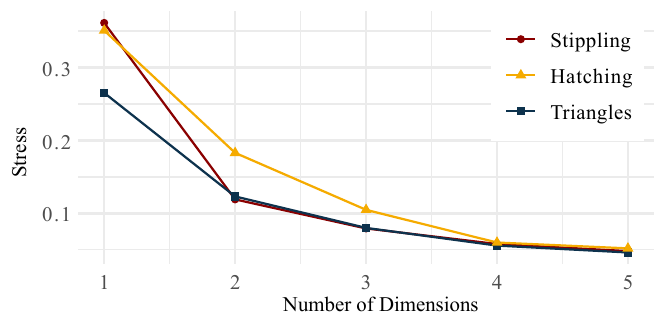}
    \caption{Scree plot for stippling, hatching, and triangles.
    Two dimensions seem to be sufficient to describe all three kinds of data.}
    \label{fig:scree}
\end{figure}
\begin{figure*}[tb]
\centering
\begin{subfigure}[t]{0.26\linewidth}
\centering
\includegraphics[width=\linewidth]{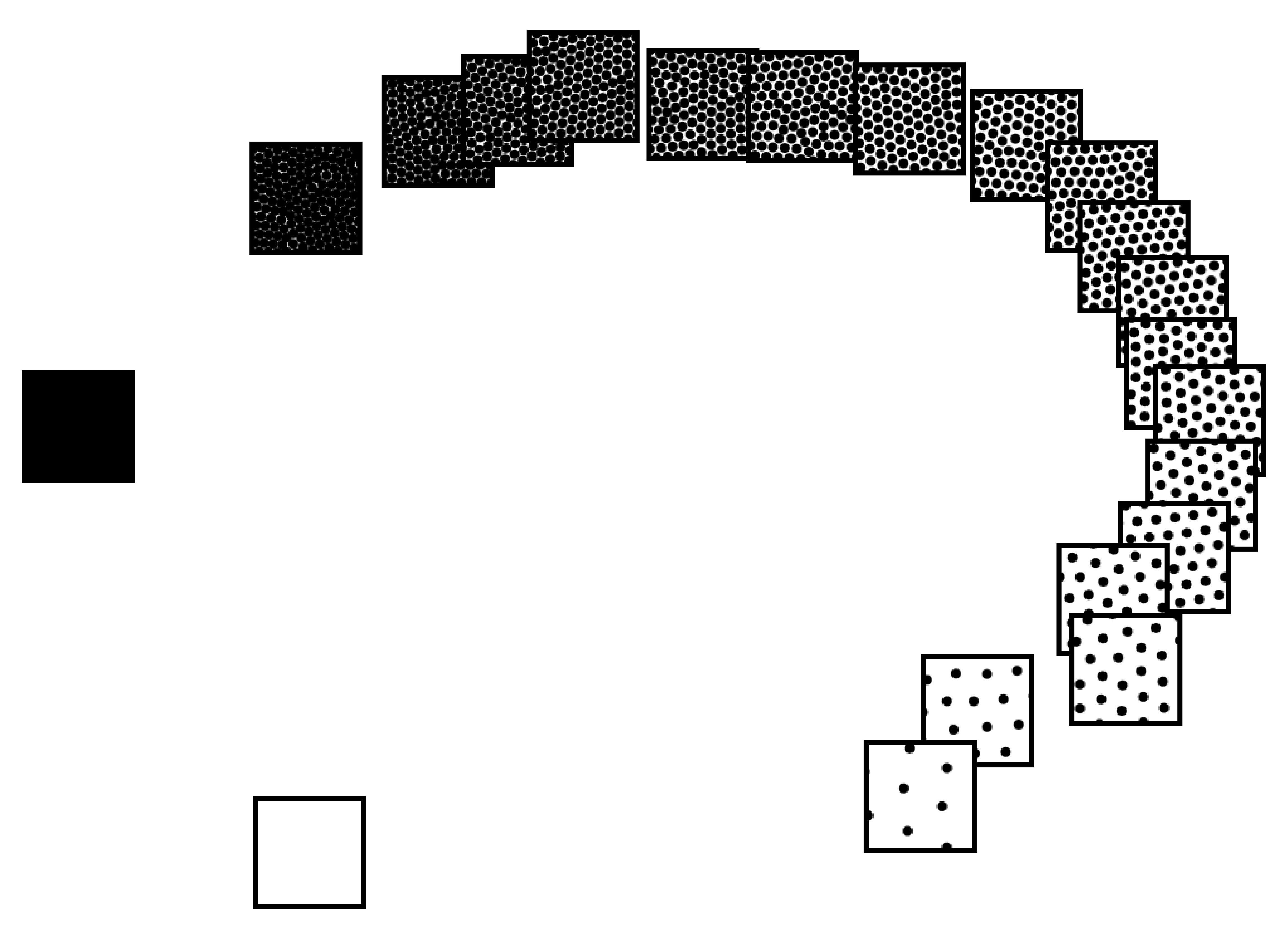}
\caption{Stippling.}
\label{fig:gspace stippling}
\end{subfigure}
\hfill
\begin{subfigure}[t]{0.23\linewidth}
\centering
\includegraphics[width=\linewidth]{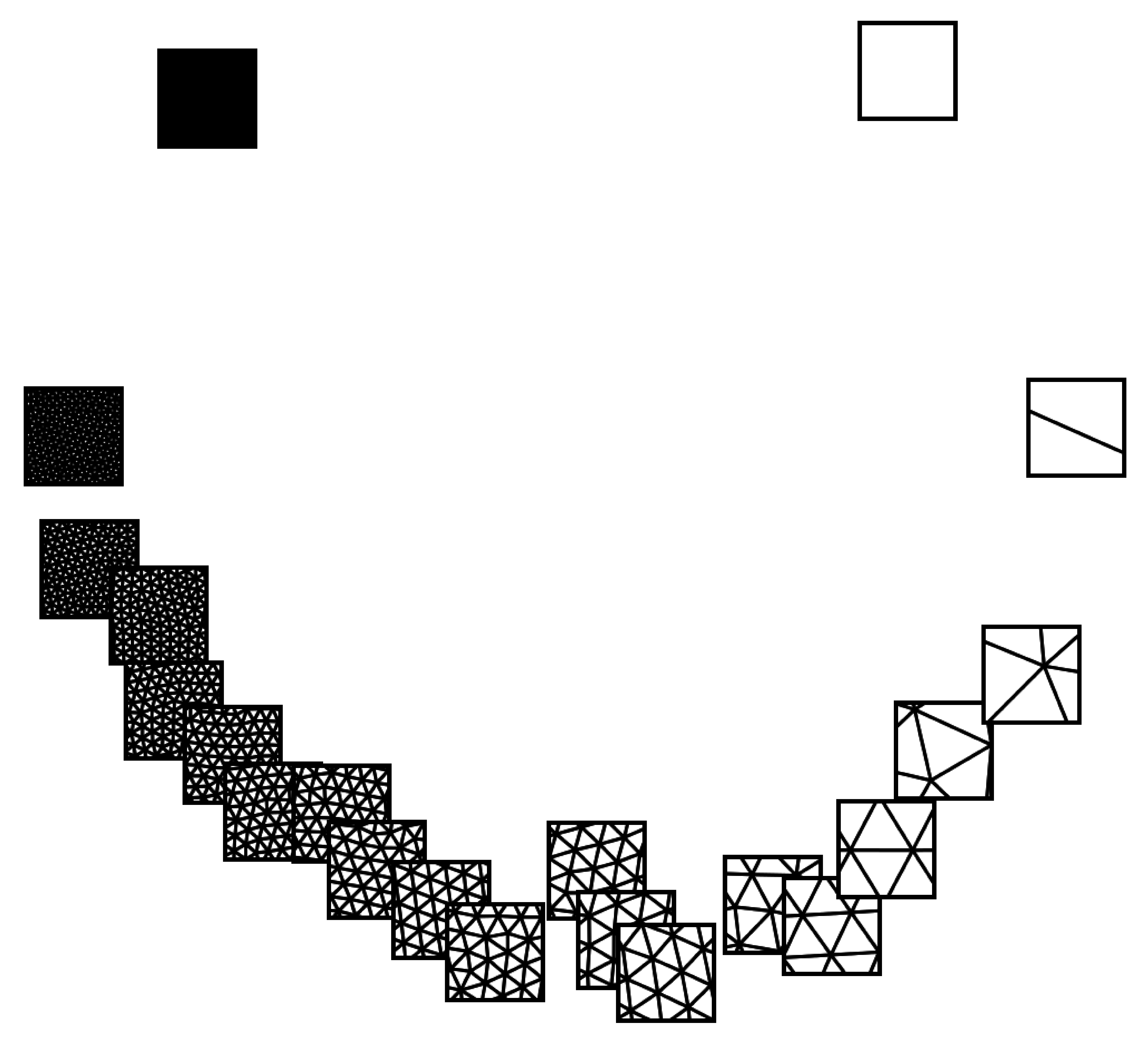}
\caption{Triangles.}
\label{fig:gspace triangles}
    \end{subfigure}
    \hfill
\begin{subfigure}[t]{0.44\linewidth}
\centering
    \includegraphics[width=\linewidth]{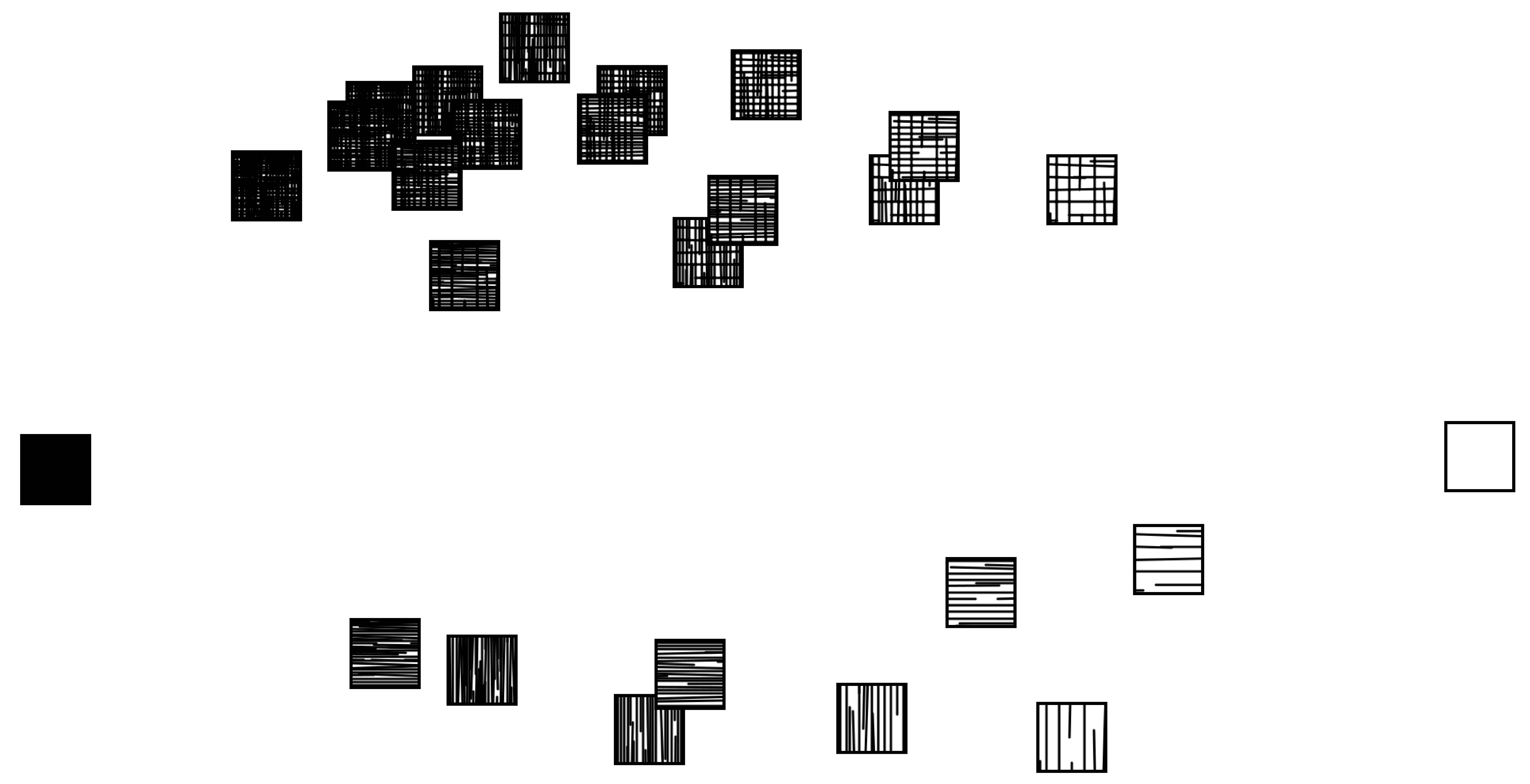}
    \caption{Hatching.}
    \label{fig:gspace hatching}
\end{subfigure}
\caption{Perceptual Spaces.}
    \end{figure*}
\begin{figure}[b]
    \centering
    \includegraphics[width=0.85\linewidth]{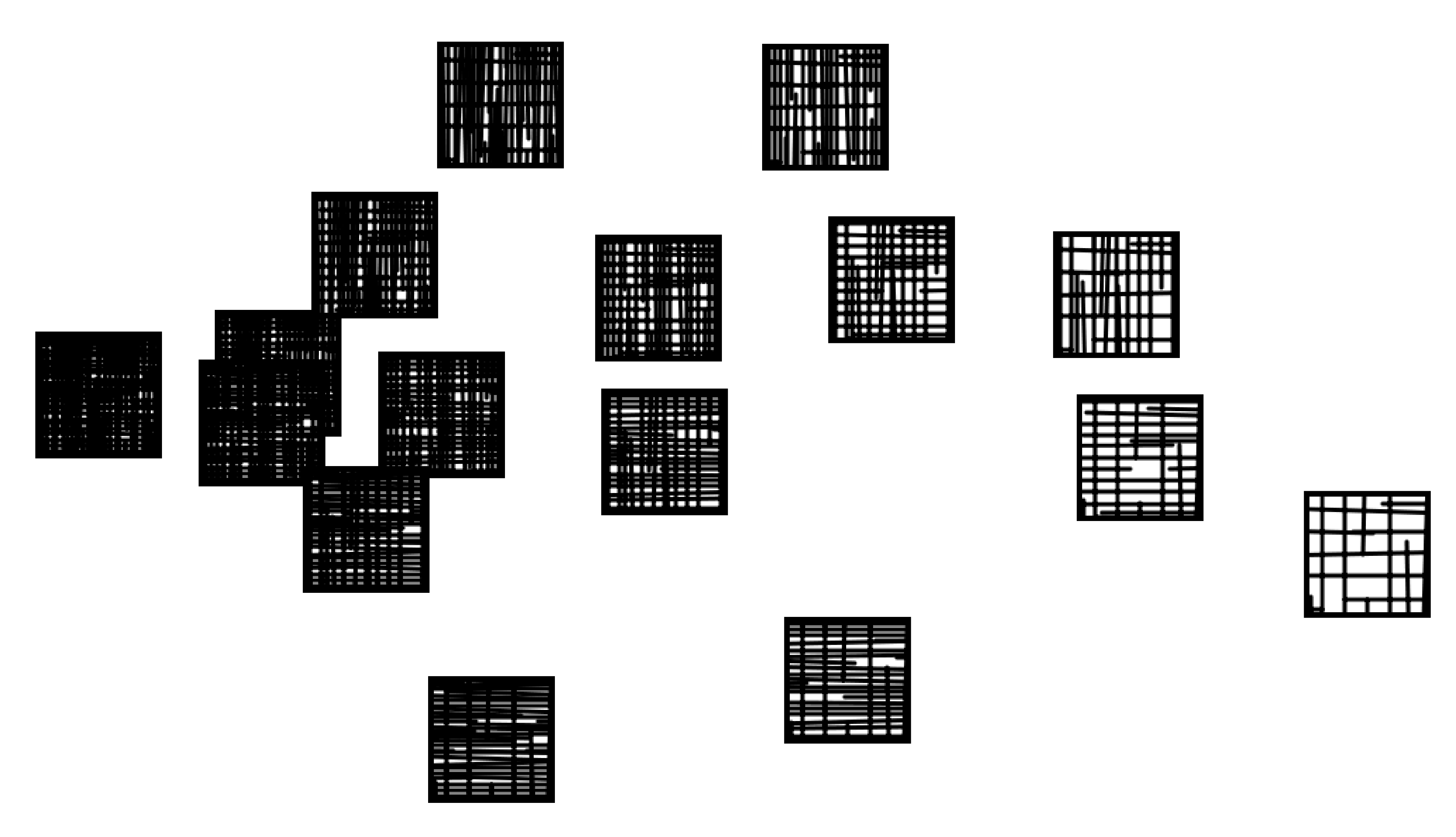}
    \caption{Perceptual Space of Crosshatching.}
    \label{fig:gspace crosshatching}
\end{figure}

\subsection{Study Procedure}
\label{sec: pe method}
We conducted three online perceptual studies, one for each of the evaluated texture types.
The setup for each study was the same.
We recruited 60 participants for the studies, 20 participants per texture type.
Each participant had to compare all possible pairs of textures per assigned texture type.
Because we assume that the results are symmetric---it does not matter if the lower density is in the right or the left image---there are $N / 2 * \left(N - 1\right) + N$ comparisons per participant.
$N$ is the number of stimuli for the relevant texture type.
$N = 21$ for stippled and triangulated textures and therefore 231 pairs had to be compared per participant.
For the hatched textures we have six density levels per dimension therefore 36 stimuli. However, eleven of these stimuli are completely covered in primitives, therefore we only need to consider $N=26$.
The 26 stimuli result in 351 comparisons per participant.
This means there were a total of $2 \cdot 20 \cdot 231 + 20 \cdot 351 = 16 260$ evaluated tasks in our studies.
We chose a medium-gray background for presenting the stimuli, as medium gray corresponds to the average density value over all stimuli.
An example task can be seen in Figure \ref{fig:example_screen}.
The participants had to rate the difference in stimuli on a scale from 1 (very similar) to 9 (very different).
\review{This question allows participants to assess the difference between the stimuli without specific instructions and thus provides an unbiased view of how they perceive textures.}

The general procedure of each study was as follows.
First, the participants were informed about the study and consented to participate.
The tasks were explained very briefly.
We gave no example images and provided no anchoring.
We chose this setup because while it may increase the noise in the data, it avoids bias.
Then, the comparison tasks began.
\review{
We used the crowdsourcing platform Prolific to recruit participants.
In comparison to-lab based experiments, crowdsourcing adds some noise to the measurements.
However, this variation also makes the experiment more realistic.
For a comparison of crowdsourced and lab-based experiments, we refer to the paper of Gadiraju et al.~\cite{Gadiraju2017}.
}
The inclusion criteria for the participants were normal or corrected to normal vision.
\review{
The participants were instructed to participate on desktop devices only to make the conditions more comparable.}
The average reward per study was at least £9 per hour.
 
\subsection{Analysis of Study Results}
\label{sec: pe analysis}
We use multidimensional scaling (MDS) to analyze our results.
MDS takes a symmetric matrix of (dis)-similarity ratings and finds a lower dimensional representation of the data that preserves the pairwise distances as accurately as possible.

First, we explored our data results visually.
The participants in our study had to rate their subjective perceived distances for all pairs---there are no objective correct or wrong answers.
However, a good indication that participants genuinely tried to answer sensibly is if the similarity ratings for stimuli with the same density level are rated as highly similar.
We only found one of the sixty participants with an apparently random answer distribution.
All other participants appeared to answer in good faith.
We excluded the results of this one participant from the analysis and recruited an additional participant instead.
All participant responses can be found in the supplemental material.

In addition to the similarity matrix, MDS requires the choice of dimensions as an input parameter.
If a suitable number of dimensions is not known beforehand, typically MDS is run for a series of input dimensions.
\review{Choosing the appropriate number of dimensions is a common challenge when using the MDS approach.}
In a \emph{scree plot}, the \emph{stress}, an optimization parameter for MDS, is plotted against the number of dimensions.
A distinct decrease in stress for a particular dimension is then interpreted as the number of dimensions suited to explain the data.
Stress values are usually normalized between 0 and 1. 
A popular implicit stress normalization is Kruskal's \emph{stress-1}~\cite{Kruskal1964}, which we will use during the analysis.
As a rule of thumb, stress values below 0.2 are often considered acceptable, however, it is important to not only consider stress values.
High-stress values do not necessarily implicate a bad goodness-of-fit but might also be a consequence of a high error in the data \cite{Borg2005}.
We ran MDS on the averaged results of all participants to determine an adequate number of dimensions to describe the perceptual spaces.
Figure \ref{fig:scree} shows a scree plot for all texture types.
For stippling and triangulation, a steep decline from one to two dimensions is visible, indicating, that two dimensions are sufficient.
For hatching, while the decline is less pronounced, the stress appears to be sufficiently low in two dimensions as well.

In the following, we use weighted MDS (INDSCAL) in which the individual dissimilarity matrices of each participant are considered.
Averaging does only produce good results if the correlation between the individual matrices is high, therefore, INDSCAL algorithms are the recommended choice for multiple participants \cite{Cunningham2011}.
Figure \ref{fig:gspace stippling} shows the perceptual space of stippling.
While one dimension was not sufficient to describe the perceptual space, the stippled textures seem to lie in a 1D manifold in 2D space.
Additionally, the perception of stippling seems to be highly non-linear for very empty (density $\approx 0$) and very dense (density $\approx 1$) regions.
The perceptual space of triangles (Figure \ref{fig:gspace triangles} is similar to the perceptual space of stippling---a 1D manifold in 2D space with larger spacing for very empty and very dense areas.

As expected, the perceptual space of hatching (Figure \ref{fig:gspace hatching}) is more complicated.
Excluding the empty and the completely covered hatched textures, the textures are divided into two groups.
One group consists of all crosshatched textures and the other group of all textures with only one hatching direction.
Because of this clustering, we would advise against using combinations of hatching and crosshatching for encoding one-dimensional data.
The perceived distances at a jump from one cluster to the other would be very high.

\review{One dimension of the perceptual space seems to encode the mean density of the textures (x-axis); however, there is no clear interpretation for the other dimension (y-axis).}
For the cluster of textures with only one hatching direction, the horizontally hatched textures have a higher y-value than the corresponding vertically hatched textures.
Figure \ref{fig:gspace crosshatching} shows the results of an MDS analysis on the subset of crosshatched textures.
Again, one dimension seems to encode the density of the textures.
Along the other dimension, the textures seem to be aligned according to the density of horizontal and vertical dashes.
Textures with dense horizontal lines have low y-values, and textures with dense vertical lines have high y-values.
\review{
While we try to interpret the relationship between the physical and the perceptual dimensions---such as \emph{similar density} for the major dimension---it is important to mention that these post hoc interpretations are hypotheses only.
Further experiments are necessary to uniquely interpret the revealed structures.
}

\begin{figure}[H]
    \centering
    \scriptsize
    \begin{subfigure}[t]{0.47\linewidth}
    \includegraphics[width=\linewidth]{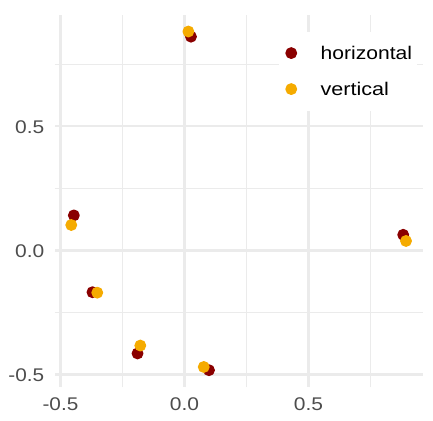}
    \caption{Horizontal and vertical hatching.}
    \label{fig:horizontal_vertical}
    \end{subfigure}
    \hfill
    \begin{subfigure}[t]{0.47\linewidth}
    \centering
    \includegraphics[width=\linewidth]{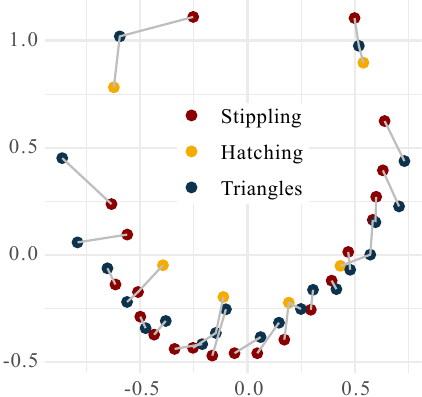}
    \caption{Stippling, 1D hatching, and triangles.
    Equal density levels are connected by lines.
    }
    \label{fig:aligned}
    \end{subfigure}
    \caption{Comparison of several perceptual spaces.}
\end{figure}

Figure \ref{fig:horizontal_vertical} shows the results of separate MDS analyses for horizontal- and vertical-only hatching.
The results of MDS are rotationally invariant; thus, we aligned the two point sets using Kabsch's algorithm \cite{Kabsch1976}, which minimizes the root mean squared deviation.
We will create uniform levels for horizontal hatching only as the results align well.

Similar to the perceptual spaces of stippling and triangles, the perceptual space of 1D hatching appears to be a 1D manifold in 2D space.
Figure \ref{fig:aligned} shows the aligned results of stippling, horizontal hatching, and triangles using Kabsch's algorithm.
Note that the x-axis for stippling is inverted in Figure \ref{fig:aligned} to match up with the perceptual spaces of horizontal hatching and triangles.
In general, the curves align relatively well, the deviations are higher for lower-density values.
We suppose that for lower-density values, the perception is more strongly related to the perception and arrangement of the individual primitives.
For higher densities, the perception is probably more closely related to the average density.
\review{Completely covered and empty textures were also perceived as relatively similar and we suspect that this contributes to the necessary 2D embedding.}

\FloatBarrier
\begin{figure}[tb]
\centering
\scriptsize
\includegraphics[width = \linewidth]{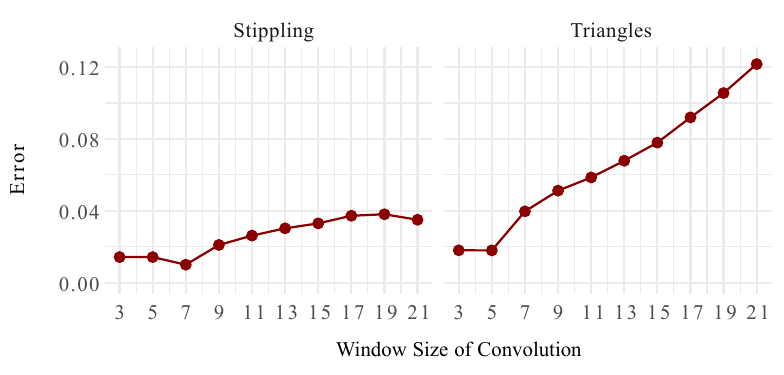}
    \caption{Errors for the curve fitting for stippling and triangle textures based on the window size.}
    \label{fig:error}
\end{figure}
 \begin{figure*}[tb]
    \centering
    \scriptsize
    \includegraphics[width= \linewidth]{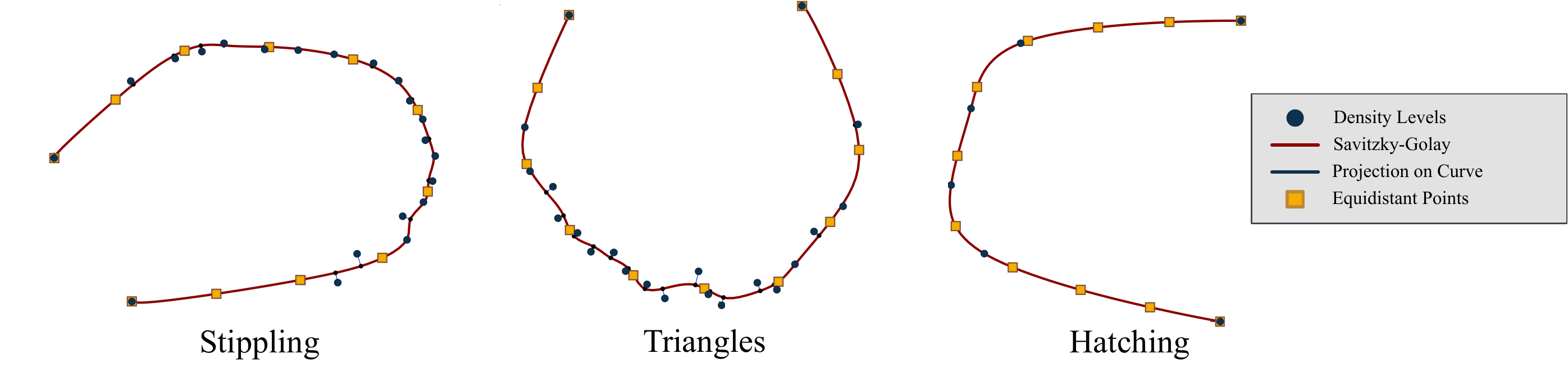}
    \caption{Reparameterization of the perceptual spaces. The density levels are approximated by a curve using a Savitzky-Golay filter.
    Then, equidistant points are sampled on the curve.
    The density values for the equidistant points are calculated by linearization between projections of neighboring density levels onto the curve.
    }
    \label{fig:perceptualUniform}
\end{figure*}
\begin{figure*}[tb]
\centering
\includegraphics[width = \linewidth]{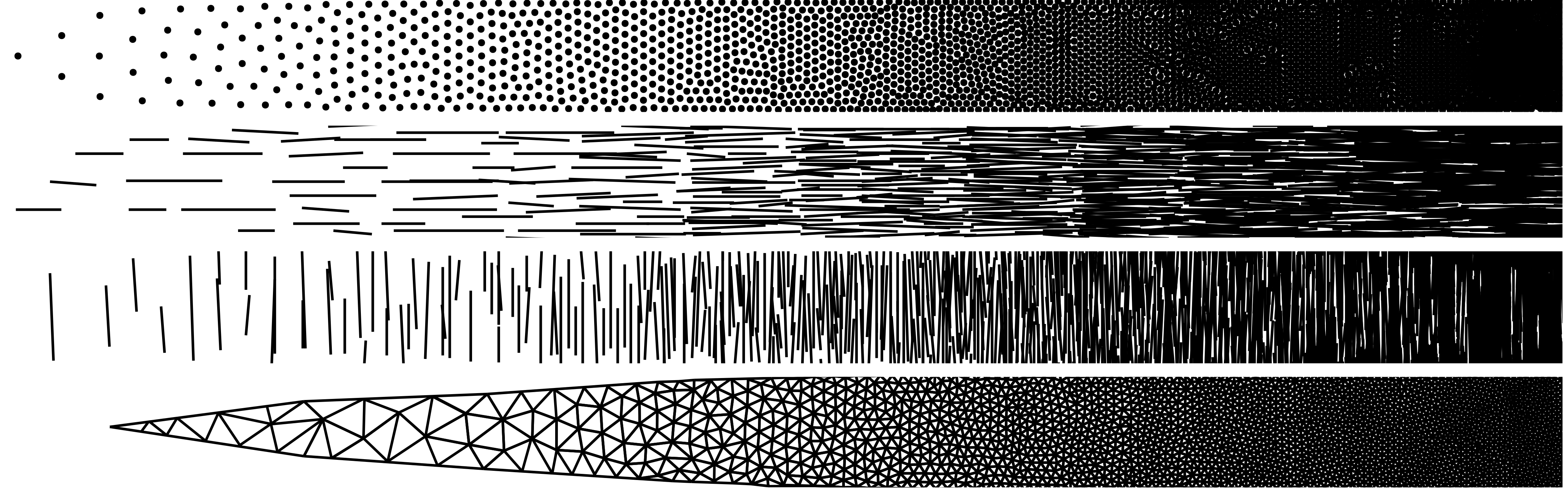}
\caption{\review{Continuous reparameterization examples. From top to bottom: stippling, horizontal hatching, vertical hatching, triangles.}}
\label{fig:continuous_textures}
\end{figure*}

\begin{figure}[!b]
\centering
\includegraphics[width = 0.85\linewidth]{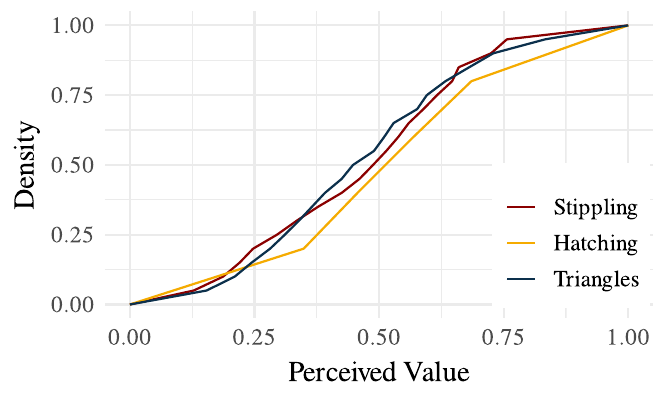}
    \caption{Reparameterizations of the perceptual spaces. For a desired perceived value, the density values can be inferred, and vice versa.}
    \label{fig:reparameterization}
\end{figure}
\section{Construction of Perceptually Uniform Textures}
\label{sec: construction}
Based on the points resulting from the MDS (INDSCAL) algorithm, we want to reparameterize the perceptual spaces to find perceptually uniform textures.
Because the perceptual spaces only locally resemble 1D Euclidean space---they are 1D manifolds in 2D space---our succession of textures will only be locally perceptually uniform.
This means that the distances between neighboring textures will be perceived similarly for all textures.
However, the results cannot be perceptually uniform on a global scale.
Given a sequence of textures $\left[a, b, c\right]$ with equal perceptual distances $d_{ab}, d_{bc}$, the distance $d_{ac}$ will not be equal to the sum $d_{ab} + d_{bc}$.

For the construction of locally perceptually uniform textures, we need to find a curve that well approximates the points.
To achieve this, we use the Savitzky-Golay filter~\cite{SavitzkyGolay1964}.
By convolving adjacent subsets of points with a low-degree polynomial, the resulting curve approximates the points.
An advantage of the Savitzky-Golay filter is that high frequencies are included in the calculation. 
This preserves the distribution of characteristics such as relative maxima and minima.
As input for the filter, the degree of the polynomial and the size of the convolution window are required.
The window size needs to be odd and in the range between the polynomial degree and the number of points.
To find a good combination of parameters, we determined the sum of the squared distances from the points to the projection on the resulting curve.
We determined the error for all window sizes with the given degree of three for the stippling and triangle textures.
For the hatching textures, we have six density levels instead of the 21 levels for the other two texture types.
We found that for this lower number of levels, a polynomial of degree two is necessary for the approximation. With a window size of five, we could achieve a perfect fit.
Figure~\ref{fig:error} shows the errors of the fitted curves for the stippling and triangle textures.
For our final curve fits, we chose the window sizes with minimal error.
The minimum for stippling is at a window size of seven, for triangles, the error is equal for a window size of three and five.
For smoother results, we used a window size of five.

Next, we chose points on the curve that are uniformly distributed.
For each texture type, we first reparametrized the curve $c$ according to its arc length.
This allows us to determine the uniform distribution of $d$ into $n$ parts of the curve by evaluating: $c(t),\,t\in\{0,L/n,2\cdot L/n\ldots,L\}$, where $L$ represents the length of the curve.
To identify the corresponding density for each point $c(t_i)$, we need to locate the neighboring projected points $c(t_l)$ and $c(t_r)$ of $c(t_i)$ such that $t_l\le t_\le t_r$.
According to the parameterization of the curve, these projected points are closest to $c(t_i)$, i.e., there are no other projected points on the curve between $[t_l,t_i]$ and $[t_i,t_r]$.
The density $d(t_i)$ of $c(t_i)$ is then determined by linearization: 
$d(t_i)=\lambda d(t_l)+(1-\lambda)d(t_r)$, with $\lambda=\frac{t_r-t_i}{t_r-t_l}$.
Figure~\ref{fig:perceptualUniform} shows the reparameterizations of the various textures with equidistant points. 

\review{
Figure \ref{fig:continuous_textures} displays examples of these reparameterizations.
In each case, we sampled 30 equidistant levels along the reparameterization curve.
For stippling and triangles, for each level, the textures were filled with the appropriate number of primitives required for the level's density.
Afterwards, we concatenated the textures and relaxed the primitives using Lloyd's algorithm \cite{Lloyd1982}. 
Due to the inherent connectivity of all triangles, it becomes challenging to sample triangle densities accurately when the density varies across the entire texture.
Since triangles cannot be disconnected, this can lead to an increase in the number of lines and, consequently, a higher overall density in parts of the texture with lower required densities. In the example, the density error is bigger than 0.05 for densities smaller than 0.1.
For hatching, we concatenated the textures first and then populated the whole texture until the required densities were met.
}\\

\begin{figure}[tb]
\centering
\scriptsize
\includegraphics[width = 0.85\linewidth]{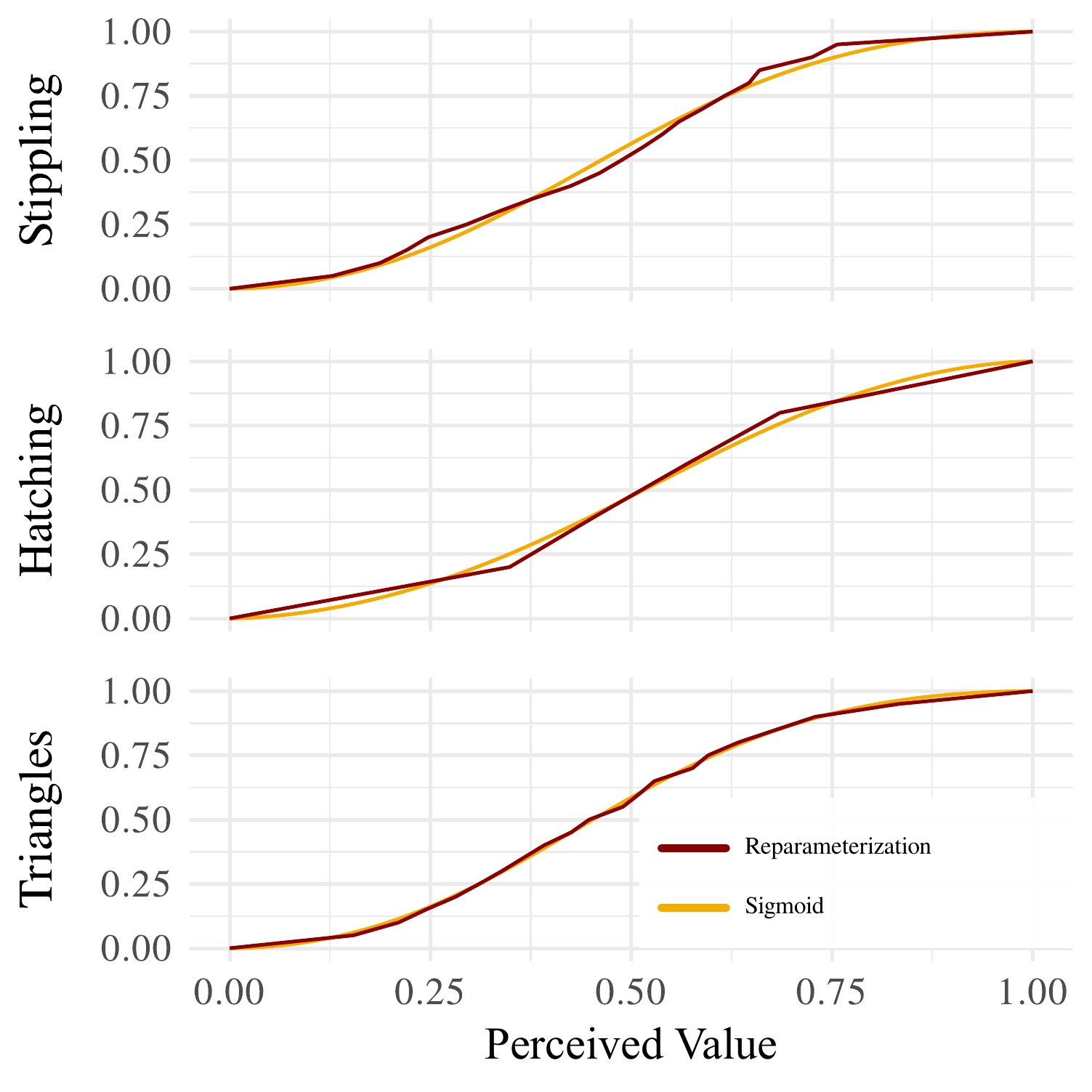}
    \caption{\review{Reparameterization functions mapping perceptual space to density space, shown alongside their corresponding sigmoid fits.}}
    \label{fig:sigmoids}
\end{figure}

\noindent\textbf{Approximation of Perceptually Uniform Textures.}
In the following, we generated 1000 uniformly distributed points on the curve (in the perceptual space) and determined their original density values for all texture types, see Figure \ref{fig:reparameterization}.
\review{
Psychometric functions---functions which describe the relationship between a physical stimulus and the perceptual response---are typically described by sigmoid functions~\cite{Wichmann2001}.
Indeed, this also seems to be the case for our data.
%
When plotting the density values against the perceived values, we noticed a similarity with a general derived sigmoid function:
\begin{equation*}
    f(x)=\frac{1}{1+(\frac{1}{a}-1)(\frac{1}{x}-1)^{b} },
\end{equation*}
We refer to the supplementary material for a rational explanation of the choice of this concrete type of sigmoid function.
}
Therefore, we approximated the points and determined the best-fit values, see Table~\ref{tab:curve_coe_2}.
\begin{table}
\centering
\caption{\review{The variables for the best fit sigmoid function with the corresponding RMSE.}
}
{\begin{tabular}{lcccc}\toprule 
Texture Type & a & b &  RMSE \\ \midrule
Stippling & 0.5644  & 1.7361 & 0.0233 \\
Hatching & 0.4753 & 1.5918 & 0.0225 \\
Triangles & 0.5859 & 1.8120 &  0.0089\\
\bottomrule
\end{tabular}}
\label{tab:curve_coe_2}
\end{table}
\begin{table}
\centering
\caption{Seven perceptually uniform density levels generated with our reparameterization.
Excluding the extrema, these density levels were used in the application examples.
}
\begin{tabular}{llllllll}\toprule
Texture Type & \multicolumn{5}{l}{Levels} \\ \midrule
Stippling & 0.083 & 0.298 & 0.523 & 0.852 & 0.966 \\
Hatching & 0.096 & 0.191 & 0.477 & 0.768 & 0.894 \\
Triangles & 0.061 & 0.290 & 0.576 & 0.836 & 0.950 \\
\bottomrule
\end{tabular}
\label{tab:levels}
\end{table}
Figure \ref{fig:sigmoids} shows the sigmoids plotted against the perceptual reparameterizations.
%
Interestingly, when determining the inverse function, we obtain an approximation of perceived values from density values.
%
The inverse function is:\review{
\begin{equation*}
    f^{-1}(x)=\frac{1}{1+\sqrt[b]{\frac{a(x-1)}{x(a-1)}}}
\end{equation*}}

\section{Application}
\label{sec: application} 
In the following section, we present some application examples.
\review{
In addition to visualizing individual scalar fields, there are scenarios where the simultaneous analysis of two scalar fields becomes crucial.
In such cases the obvious choice of color-encoding is not sufficient for displaying all relevant data, another visual channel needs to be used for the second scalar field.
One such example is the previously mentioned work by Meuschke et al.~\cite{Meuschke_2017_TVCG}.
They used a combination of color-encoding and image-based hatching to visualize two scalar fields simultaneously on aneurysm walls.
For the assessment of rupture risk in cerebral aneurysms the understanding of the correlations between hemodynamic factors and wall-related characteristics, like wall shear stress and wall thickness, is essential.
By visualizing these fields simultaneously, it becomes possible to identify spatial relationships between regions of high-shear stress and thin walls.
}

\review{
In the following section, we show additional application scenarios.
}
For each application scenario, we use surface meshes to which we apply the textures to encode information.
First, we computed a global parameterization of each surface mesh.
This provides global texture coordinates that can be used for texturing.
The parameterization is computed using the as-rigid-as-possible (ARAP) method, which tries to keep the area of the triangles constant under the parameterization~\cite{Liu2008}.
With the given texture coordinates we applied the textures to the surface mesh.
Depending on the underlying scalar field, the textures will vary in tone.
\review{In many application scenarios, the data is categorized into discrete bins, e.g., low, medium, or high risk.
In other scenarios, the displayed data should be varied continuously, e.g., distance to another surface.
In the following, we show examples of both types of visualization.
}

\begin{figure}[tbh]
    \centering
    \includegraphics[width=\linewidth]{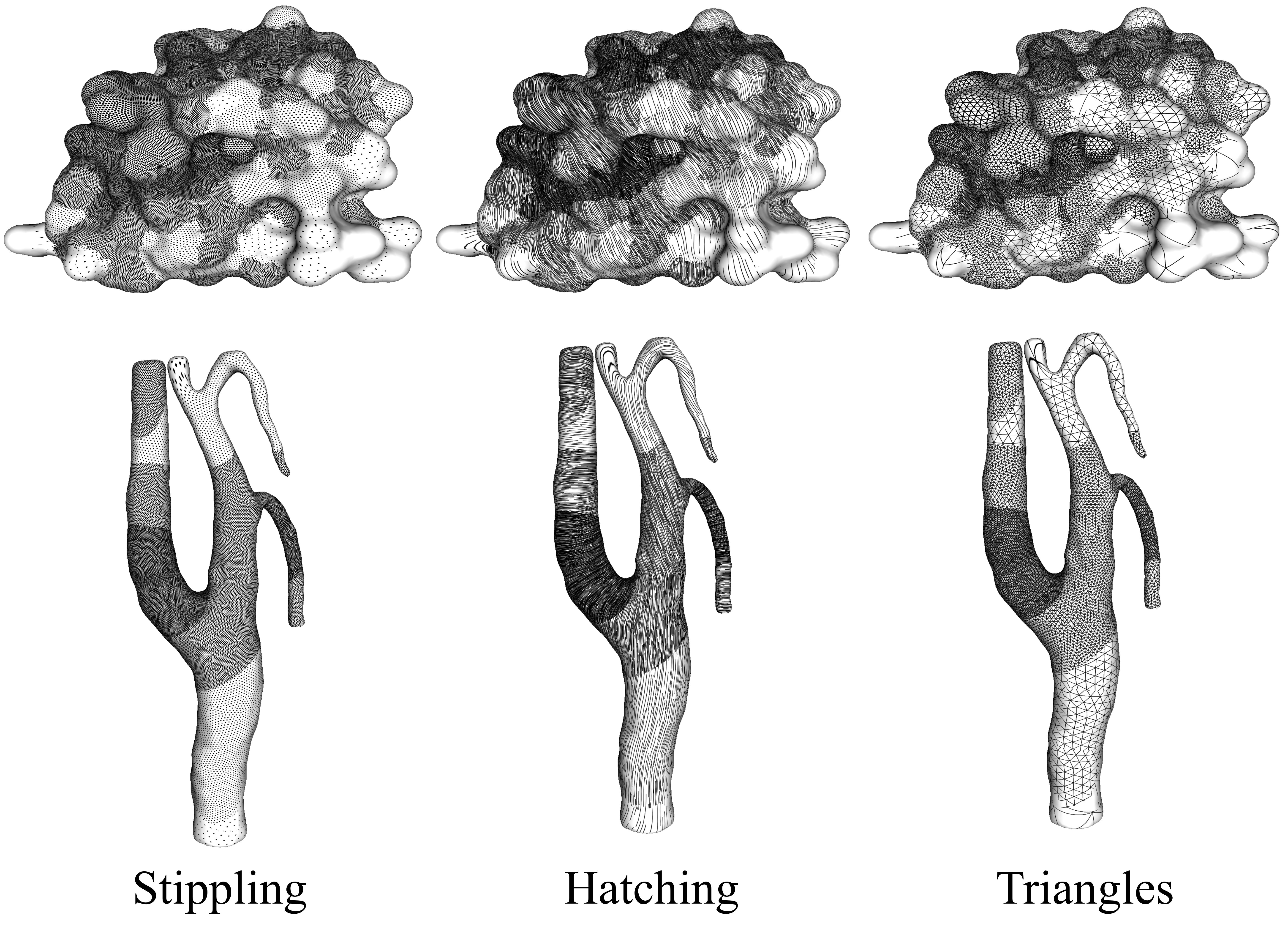}
    \caption{\review{Application examples for perceptually uniform illustrative textures with discrete data values. The first row shows a molecular surface, the second row displays a carotid artery model.}}
    \label{fig:application examples}
\end{figure}

\review{
For the discrete data values, we chose five perceptually uniform textures based on our reparameterizations.
First, we sampled seven equidistant points on the curves.
We then excluded the white and the black texture to arrive at our desired five density levels which can be found in Table \ref{tab:levels}.
All vertices are divided into five bins based on the underlying scalar field.
For each bin, a different tone texture is selected and this texture is then sampled with the corresponding texture coordinates of the parameterization.
Preexisting illustrative textures can be used for each density value.
Each vertex and interpolated fragment receives the corresponding color value from the underlying tone texture.
This method can lead to sharp transitions between areas of different densities.
}
\review{
Figure \ref{fig:application examples} shows two application examples, a molecular surface, and a carotid artery model.
}
Molecular visualization is important for understanding the structure, behavior, and interactions of molecules and for communicating scientific concepts to a broader audience.
For these purposes, many different molecular properties are visualized on molecular surface models.
In our application example, we display the molecule's b-factor.
The b-factor describes the thermal motion of the molecule's atoms.
It is often considered a measure of positional uncertainty.
\review{
While uncertainty visualization for molecular data enhances the accuracy and reliability of visualizations, usually other molecular properties need to be displayed simultaneously.
Therefore, data such as these particularly benefit from our illustrative textures.
The color channel stays open for the primary attribute while the uncertainty can be displayed as the texture pattern.
}
The molecule in Figure \ref{fig:application examples} is human insulin \cite{Timofeev2010, Timofeev2010a}.\\
\FloatBarrier
 \begin{figure}[tbh]
    \centering
    \includegraphics[width=\linewidth]{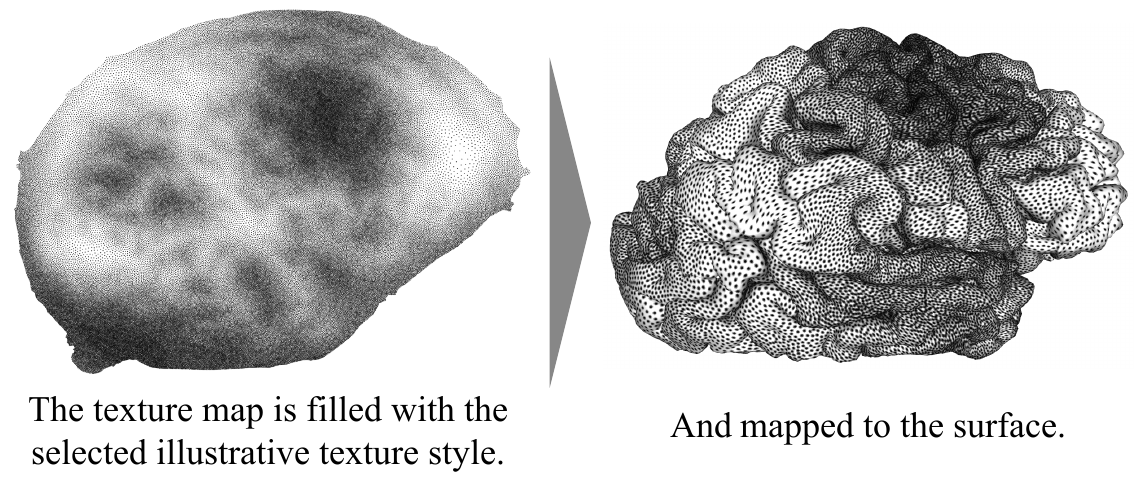}
    \caption{\review{Continuous distance representation on a brain model.}}
    \label{fig:application continuous}
\end{figure}
\review{
The other application example concerns medical visualization.
}
Here, we used a carotid data set~\cite{eulzer_pepe_2023}, parameterized as proposed in the work of Eulzer et al.~\cite{Eulzer_2021_VCBM}.
For carotid data sets, the illustrative textures can be used to show the wall shear stress or other hemodynamic properties.
These properties may indicate the risk of a stroke.
The scalarfield displayed in Figure \ref{fig:application examples} is generated artifically for illustrative purposes.\\
Note, we did not align the hatching textures along the principal curvature direction, which is conventionally done.
In our examples, we display our density levels without this constraint.

\review{Besides discrete transitions, we also show a continuous example, using a stippled visualization of a brain, see Figure~\ref{fig:application continuous}.
Medical researchers and clinicians rely on important aspects such as the amount of brain damage, anatomical location, shape, and changes to understand the temporal patterns of multiple sclerosis (MS).
For lesion analysis and exploration, it is important to encode the distance to the brain surface~\cite{Sugathan_2021_VCBM, Sugathan2022}.
In Figure \ref{fig:application continuous}, a distance field is visualized on a brain model.
In the first step, the distance values on the texture map needed to be scaled and transformed with our reparameterization derived in Section \ref{sec: construction}.
Then, the primitives need to be correctly distributed to match the density values.
We used weighted Linde-Buzo-Gray stippling~\cite{Deussen2017} in this step.
This distribution algorithm could be adapted for triangles and for hatching the method described in Section \ref{sec: pe stimuli} can be adapted.
Finally, the texture can be mapped to the 3D model.
}

\section{Discussion and Future Work}
We found that the perceptual spaces of illustrative textures, more specifically stippling, hatching, and triangles, can be successfully embedded in 2D Euclidean space.
Stippling, one-dimensional hatching, and triangles can even be described by 1D manifolds.
We found that our sample points were more spaced out at both ends of the density range.
This indicates that the perception of the textures is nonlinear in these areas.
Thus our results confirm the conclusion of Görtler et al.~\cite{Gortler2019} who investigated stippling by computing just-noticeable differences (JND).
However, they recommend a root or logarithmically scaled density mapping.
We cannot recommend this based on our results.
%
As discussed in Section \ref{sec: construction}, a sigmoid function seems to be a sensible mapping.
%
%
The perceived distances got larger on both ends of the density range.
One of the reasons for this mapping recommendation by Görtler et al. was a ``bump'' artifact in their data at a density of 0.6.
We could not confirm the existence of such a ``bump'', our stippling results formed a relatively smooth curve.
They attributed it to variations in display resolution, rasterization, and viewing distance.
Based on our results we also suspect that this bump artifact might indeed have been noise and should therefore not be used as a reason for a coarser spacing at higher densities.

\review{Our constraint-free MDS approach allowed us to discover unexpected perceptual structures: 2D embeddings for stippling, hatching, and triangles.
Our question formulation of differences instead of a more specific formulation such as more/less density provided an unbiased view of the perception of illustrative textures.
Now, in the next step, the newly revealed 2D structure needs to be further investigated with standard psychophysical procedures to determine the relationship between physical dimensions and the dimensions of the perceptual space.
Our post hoc interpretation of the mayor dimension as \emph{similar density} is a hypothesis.
While we believe that it is a sensible interpretation, future work is needed to explore this hypothesis empirically. Future work will also try to determine the meaning of the second dimension. 
Ultimately, this should refine the reparameterizations and thus generate illustrative textures that are even more perceptually uniform.
One example is completely covered and empty (black and white) textures.
They were perceived as relatively similar.
This does probably contribute to the necessary 2D embedding.
In future work, another study could be conducted in which we explicitly ask for a rating concerning the similarity in density.
This could potentially make our results more one-dimensional.
}
Utilizing the 2D embeddings of our data, we reparameterized the perceptual spaces.
The reparameterizations were then used to construct perceptually uniform textures for data encoding.
These perceptually uniform textures make visualizations more intuitive and accessible.
However, our results have some limitations.
While our texture levels all have similar perceived distances to their neighbors, the perceptual spaces prohibit globally uniform texture levels.
Because of the curved nature of the perceptual spaces---they are 1D manifolds, not 1D Euclidean spaces---the perceived differences between, e.g., density level 0 and level 0.5 is smaller than the summed perceived differences between 0 and 0.25, and 0.25 and 0.5.

Additionally, our sampling on the extreme ends of the density range could be refined. 
In future work, it would be interesting to analyze more levels between densities of 0 and 0.05 and between 1 and 0.95.
Additionally, an evaluation for 1D hatching without the crosshatches could be conducted with a finer sampling of density levels.

The perceptual space for all combinations of vertical and horizontal hatching lines was complex.
We found two distinct clusters that separate crosshatched and one-dimensional hatching textures.
Because of the high perceptual distances between these clusters, we recommend against a combined use for the encoding of one type of scalar value.
In future work, we plan to investigate other uses of crosshatching.
Different line directions have been used to encode different properties before \cite{Sajadi2013}.
Our evaluations have shown that textures with similar densities are perceived as relatively similar regardless of the hatching direction.
It would therefore be interesting to determine whether different hatching directions can nevertheless be used to encode and retrieve several values.

In future work,  different texture patterns, such as the Brodatz textures \cite{uisBrodatzTextures}, could be analyzed for information encoding.
\review{Additionally, other texture types that are more symmetric could be developed.
There, the method of determining fore- and background pixels could be more balanced, maybe leading to a more linear relationship between physical density and perceived similarity.
In general, the effect of clustering in a texture would be an interesting avenue for future work.
}

\review{We applied our derived textures to 3D surfaces to show three application examples.
However, our study focused on evaluating and mapping texture similarity in 2D.
Effects and artifacts that appear because of depth perception or properties of 3D surfaces such as curvature, could interfere with the perception of the textures.
A more comprehensive investigation of texture perception on 3D surfaces is a topic that requires further in-depth research.
Related to depth perception, a sensible generation method for mipmaps for the triangle textures needs to be developed.
}
For stippling and hatching, tonal art maps~\cite{Praun2001} can be used.

\review{We presented two discrete and one continuous application example.
Continuous triangulated texture maps are problematic if small density values are present on the surface.
The connectivity of the individual triangles prevents very small density values.
Other approaches, such as disconnected individual triangles need to be developed to be able to accurately display these small values in continuous texture maps.}

Similar to the work by Görtler et al.~\cite{Gortler2019}, the JND for Hatching and Triangles could be determined as an extension of our work.
In most visualization scenarios, we aim to avoid using barely distinguishable textures.
The perceived distances need to be larger for typical visualization tasks.
However, the JND would be interesting as an upper boundary for the number of textures that should be used in visualizations.

\review{
Illustrative textures can serve as an alternative to colormaps, or they can be used in combination with colormaps for enhanced visual encoding.
In future work, we want to analyze the combined perceptual space and determine whether information can be retrieved accurately if different data is encoded in color and the density level of illustrative textures.
Additionally, other parameters could influence the perceptual spaces.
One of the most obvious factors would be the stimulus size.
Görtler et al.~\cite{Gortler2019} evaluated the JND for several stipple sizes.
They found differences in the linearity of the perception for different stipple sizes.
The perception for stipple sizes of 8---the size we used for our evaluations---was closest to linear.
The effects of primitive size on perceptual spaces need to be evaluated for all texture types.
}
\review{
Related is the work by Acevedo et al.~\cite{Acevedo2006} who evaluated perceptual interactions between brightness, size, and spacing in parameterized 2D icon-based visualization methods.
They found that icon brightness outperforms spacing and size in their metrics \emph{spatial resolution} and \emph{number of data values displayable at each point}.
While these results might not be directly applicable to illustrative textures, they could indicate that brightness and in turn, possibly color should be used for a primary attribute and illustrative encoding for the secondary attribute.
However, more specific verification of interactions for illustrative textures is necessary.
In our experiments, we found the basic metric form of the perceptual space for the illustrative textures, now further experiments are necessary to explore the parameters and perceptual effects---such as contrast ratios, gestalt figure perception, glare effects, penumbra and other edge-falloffs for shadows and bloom, etc.---that could modulate this perceptual space.
}

\section{Summary and Conclusion}
We explored the perceptual spaces of illustrative textures to determine perceptually uniform texture levels.
First, we constructed a succession of textures for stippling, hatching, and triangles.
The textures for stippling and triangulation varied along one dimension: The density of the visual primitives.
We sampled the whole range of densities linearly from empty, to completely covered.
For hatching, the textures consist of two dimensions.
Vertical lines and horizontal lines.
We sampled the density for each dimension independently.
Our final hatching texture space was created by overlaying every possible combination of texture pairs with different hatching directions.

To determine the perceptual spaces that correspond to our texture spaces, we conducted three crowdsourced studies.
In each study, one of the texture types was analyzed.
During the study, the participants had to rate the differences between pairs of textures.
Each density level was compared to every other density level and to itself.
Using the results from the studies, we constructed perceptual spaces using multidimensional scaling (MDS).
We found that all spaces can be adequately described by two dimensions.
For stippling, 1D hatching, and triangles the perceptual spaces seem to be 1D manifolds in 2D space, that is, they locally resemble 1D Euclidean space.
The perceptual distances between our linearly sampled textures are bigger on the extreme ends of the density range.
This means that for intuitive encodings of data, the very small and the very large densities need a more fine-grained sampling.
When taking crosshatching into account the perceptual space for hatching is more complex.
Crosshatched textures are perceived as very different from textures that are only hatched along one dimension.
\review{
Apart from this distinction, most of the differences seem to be explained by one dimension.
}

Using our results from the MDS analysis, we reparameterized the perceptual spaces of stippling, horizontal hatching, and triangles.
We exploited the structure of the perceptual spaces to fit curves to the data.
The curves were then uniformly sampled. 
These equally spaced sample points correspond to perceptually uniform levels.
We determined the corresponding density levels by linearization between the original density levels.
The reparameterizations are well described by sigmoid functions.
%
Therefore, we recommend sigmoids as suitable density mappings for perceptually uniform mappings from perceptual space to density space.
%
Finally, we showed application examples for our density levels.
\review{
We chose application examples from biomedical visualization, where often multiple scalar fields need to be displayed simultaneously.
However, the density levels can be applied to any type of one-dimensional data on surfaces.
When using illustrative textures for one attribute of the data, the color channel remains free for visualizing another attribute.
}

To conclude, we explored the perceptual spaces of illustrative textures.
We found that stippling, 1D hatching, and triangles can be described by 1D manifolds in 2D space.
We showed that the perceived distances between crosshatched textures and textures hatched in only one direction are large.
Therefore, we would advise against using a combination of 1D hatching and crosshatching for encoding 1D data.
We reparameterized the perceptual spaces to provide perceptually uniform texture levels for information encoding.
%
For an approximated perceptually uniform mapping, we recommend sigmoid functions for mappings from perceptual values to density values.
%
This density mapping should make visualizations more intuitive and accessible.

\acknowledgments{%
This work was partially funded by the Deutsche Forschungsgemeinschaft (DFG, German Research Foundation) –- Project-ID 437702916.
}

\bibliographystyle{abbrv-doi-hyperref}

\bibliography{Vis2023}

@article{Wichmann2001,
  doi = {10.3758/bf03194544},
  url = {https://doi.org/10.3758/bf03194544},
  year = {2001},
  publisher = {Springer},
  volume = {63},
  number = {8},
  pages = {1293--1313},
  author = {Felix A. Wichmann and N. Jeremy Hill},
  title = {{The Psychometric Function: I. Fitting,  Sampling,  and Goodness of Fit}},
  journal = {Perception \& Psychophysics}
}

@incollection{Gadiraju2017,
  doi = {10.1007/978-3-319-66435-4_2},
  url = {https://doi.org/10.1007/978-3-319-66435-4_2},
  year = {2017},
  publisher = {Springer},
  pages = {6--26},
  author = {Ujwal Gadiraju and Sebastian M\"{o}ller and Martin N\"{o}llenburg and Dietmar Saupe and Sebastian Egger-Lampl and Daniel Archambault and Brian Fisher},
  title = {{Crowdsourcing Versus the Laboratory: Towards Human-Centered Experiments Using the Crowd}},
  booktitle = {Evaluation in the Crowd. Crowdsourcing and Human-Centered Experiments}
}

@book{Borg2005,
  doi = {10.1007/0-387-28981-x},
  title = {Modern {{Multidimensional Scaling}}: {{Theory}} and {{Applications}}},
  shorttitle = {Modern {{Multidimensional Scaling}}},
  author = {Borg, I. and Groenen, P. J. F.},
  year = {2005},
  publisher = {{Springer}},
  googlebooks = {duTODldZzRcC},
  isbn = {978-0-387-25150-9},
  langid = {english}
}

@inproceedings{Coninx2011,
  title = {{Visualization of Uncertain Scalar Data Fields Using Color Scales and Perceptually Adapted Noise}},
  booktitle = {Proceedings of the {{ACM SIGGRAPH Symposium}} on {{Applied Perception}} in {{Graphics}} and {{Visualization}}},
  author = {Coninx, Alexandre and Bonneau, Georges-Pierre and Droulez, Jacques and Thibault, Guillaume},
  year = {2011},
  pages = {59--66},
  publisher = {{Association for Computing Machinery}},
  address = {{New York, NY, USA}},
  doi = {10.1145/2077451.2077462},
  urldate = {2022-11-30},
  isbn = {978-1-4503-0889-2},
}

@article{Deussen2017,
author = {Deussen, Oliver and Spicker, Marc and Zheng, Qian},
title = {{Weighted Linde-Buzo-Gray Stippling}},
year = {2017},
issue_date = {December 2017},
address = {New York, NY, USA},
volume = {36},
number = {6},
issn = {0730-0301},
url = {https://doi.org/10.1145/3130800.3130819},
doi = {10.1145/3130800.3130819},
journal = {ACM Transactions on Graphics},
articleno = {233},
numpages = {12},
pages = {233:1--233:12},
publisher = {Association for Computing Machinery}
}

@article{Sterzik2023,
title = {{Enhancing Molecular Visualization: Perceptual Evaluation of Line Variables with Application to Uncertainty Visualization}},
author = {Anna Sterzik and Nils Lichtenberg and Michael Krone and Daniel Baum and Douglas W. Cunningham and Kai Lawonn},
journal = {Computers \& Graphics},
year = {2023},
doi = {https://doi.org/10.1016/j.cag.2023.06.006},
pages = {401--413}
}

@article{SterzikLines,
title = {{Perception of Line Attributes for Visualization}},
author = {Anna Sterzik and Nils Lichtenberg and Jana Wilms and Michael Krone and Douglas W. Cunningham and Kai Lawonn},
  journal = {IEEE Transactions on Visualization and Computer Graphics},
note = {To appear}
}

@article{Gortler2019,
  title = {Stippling of {{2D Scalar Fields}}},
  author = {G{\"o}rtler, Jochen and Spicker, Marc and Schulz, Christoph and Weiskopf, Daniel and Deussen, Oliver},
  year = {2019},
  journal = {IEEE Transactions on Visualization and Computer Graphics},
  volume = {25},
  number = {6},
  pages = {2193--2204},
  issn = {1941-0506},
  doi = {10.1109/TVCG.2019.2903945},
}

@article{Kruskal1964,
  title = {{Multidimensional Scaling by Optimizing Goodness of Fit to a Nonmetric Hypothesis}},
  author = {Kruskal, J. B.},
  year = {1964},
  journal = {Psychometrika},
  volume = {29},
  number = {1},
  pages = {1--27},
  issn = {1860-0980},
  doi = {10.1007/BF02289565},
  urldate = {2023-03-22},
  langid = {english},
}

@article{Lawonn2018,
  title = {A {{Survey}} of {{Surface-Based Illustrative Rendering}} for {{Visualization}}},
  author = {Lawonn, Kai and Viola, Ivan and Preim, Bernhard and Isenberg, Tobias},
  year = {2018},
  journal = {Computer Graphics Forum},
  volume = {37},
  number = {6},
  pages = {205--234},
  issn = {1467-8659},
  doi = {10.1111/cgf.13322},
  urldate = {2022-06-27},
}

@inproceedings{Praun2001,
  title = {{Real-Time Hatching}},
  booktitle = {Proceedings of the 28th Annual Conference on Computer Graphics and Interactive Techniques},
  author = {Praun, Emil and Hoppe, Hugues and Webb, Matthew and Finkelstein, Adam},
  year = {2001},
  pages = {581--586},
  publisher = {{Association for Computing Machinery}},
  address = {{New York, NY, USA}},
  doi = {10.1145/383259.383328},
  isbn = {1-58113-374-X},
}

@inproceedings{Secord2002,
  title = {{Weighted {{Voronoi}} Stippling}},
  booktitle = {Proceedings of the 2nd International Symposium on {{Non-photorealistic}} Animation and Rendering},
  author = {Secord, Adrian},
  year = {2002},
  pages = {37--43},
  publisher = {{Association for Computing Machinery}},
  address = {{New York, NY, USA}},
  doi = {10.1145/508530.508537},
  urldate = {2023-03-01},
  isbn = {978-1-58113-494-0},
}

@inproceedings{Spicker2017,
  title = {{Quantifying Visual Abstraction Quality for Stipple Drawings}},
  booktitle = {Proceedings of the {{Symposium}} on {{Non-Photorealistic Animation}} and {{Rendering}}},
  author = {Spicker, Marc and Hahn, Franz and Lindemeier, Thomas and Saupe, Dietmar and Deussen, Oliver},
  year = {2017},
  pages = {8:1--8:10},
  publisher = {{Association for Computing Machinery}},
  address = {{New York, NY, USA}},
  doi = {10.1145/3092919.3092923}
}

@inproceedings{Sterzik2022,
booktitle = {Eurographics Workshop on Visual Computing for Biology and Medicine},
title = {{Perceptual Evaluation of Common Line Variables for Displaying Uncertainty on Molecular Surfaces}},
author = {Sterzik, Anna and Lichtenberg, Nils and Krone, Michael and Cunningham, Douglas W. and Lawonn, Kai},
year = {2022},
publisher = {Eurographics Association},
ISSN = {2070-5786},
ISBN = {978-3-03868-177-9},
DOI = {10.2312/vcbm.20221186},
pages = {41--51}
}

@article{Stoppel2019,
  title = {{{LinesLab}}: {{A Flexible Low-Cost Approach}} for the {{Generation}} of {{Physical Monochrome Art}}},
  shorttitle = {{{LinesLab}}},
  author = {Stoppel, S. and Bruckner, S.},
  year = {2019},
  journal = {Computer Graphics Forum},
  volume = {38},
  number = {6},
  pages = {110--124},
  issn = {1467-8659},
  doi = {10.1111/cgf.13609},
  urldate = {2022-11-29},
  langid = {english},
}

@article{Lawonn2017,
author = {Lawonn, Kai and Luz, Maria and Hansen, Christian},
title = {{Improving Spatial Perception of Vascular Models Using Supporting Anchors and Illustrative Visualization}},
journal = {Computers \& Graphics},
issue_date = {April 2017},
volume = {63},
year = {2017},
issn = {0097-8493},
pages = {37--49},
numpages = {13},
url = {https://doi.org/10.1016/j.cag.2017.02.002},
doi = {10.1016/j.cag.2017.02.002},
acmid = {3067940},
publisher = {Pergamon Press, Inc.},
address = {Elmsford, NY, USA},
}

@Article{Lawonn_2014_CGFb,
  Title   = {{Adaptive Surface Visualization of Vessels with Animated Blood Flow}},
  Author  = {Kai Lawonn and Rocco Gasteiger and Bernhard Preim},
  Journal = {{Computer Graphics Forum}},
  Year    = {2014},
  Pages   = {16--27},
  Volume  = {33(8)},
  doi = {10.1111/cgf.12355}
}

@article{Lawonn2019_Tri,
author = {Lawonn, Kai and Günther, Tobias},
title = {{Stylized Image Triangulation}},
journal = {Computer Graphics Forum},
volume = {38},
number = {1},
pages = {221--234},
doi = {https://doi.org/10.1111/cgf.13526},
url = {https://onlinelibrary.wiley.com/doi/abs/10.1111/cgf.13526},
eprint = {https://onlinelibrary.wiley.com/doi/pdf/10.1111/cgf.13526},
year = {2019}
}

@article{Xiao2022,
author = {Xiao, Yanyang and Cao, Juan and Chen, Zhonggui},
title = {{Image Representation on Curved Optimal Triangulation}},
journal = {Computer Graphics Forum},
volume = {41},
number = {6},
pages = {23--36},
doi = {https://doi.org/10.1111/cgf.14495},
url = {https://onlinelibrary.wiley.com/doi/abs/10.1111/cgf.14495},
eprint = {https://onlinelibrary.wiley.com/doi/pdf/10.1111/cgf.14495},
year = {2022}
}

@article{SavitzkyGolay1964,
  author = {Savitzky, Abraham and Golay, M. J. E.},
  booktitle = {Analytical Chemistry},
  doi = {10.1021/ac60214a047},
  journal = {Analytical Chemistry},
  number = 8,
  pages = {1627--1639},
  publisher = {American Chemical Society},
  title = {{Smoothing and Differentiation of Data by Simplified Least Squares Procedures.}},
  url = {http://dx.doi.org/10.1021/ac60214a047},
  volume = 36,
  year = 1964
}

@misc{eulzer_pepe_2023,
  author       = {Eulzer, Pepe and
                  Lawonn, Kai},
  title        = {{A Dataset of Reconstructed Carotid Bifurcation 
                   Lumen and Plaque Models with Centerline Tree}},
  year         = 2023,
  publisher    = {Zenodo},
  version      = {1.0.0},
  doi          = {10.5281/zenodo.7634644},
  url          = {https://doi.org/10.5281/zenodo.7634644}
}

@inproceedings{Eulzer_2021_VCBM,
author = {Pepe Eulzer and Kevin Richter and Monique Meuschke and Anna Hundertmark and Kai Lawonn},
title = {{Automatic Cutting and Flattening of Carotid Artery Geometries}},
journal = {Eurographics Workshop on Visual Computing for Biology and Medicine},
year = {2021},
booktitle = {Eurographics Workshop on Visual Computing for Biology and Medicine},
publisher = {Eurographics Association},
DOI = {10.2312/vcbm.20211347},
pages = {79--89}
}

@article{Sugathan2022,
title = {{Longitudinal Visualization for Exploratory Analysis of Multiple Sclerosis Lesions}},
author = {Sherin Sugathan and Hauke Bartsch and Frank Riemer and Renate Grüner and Kai Lawonn and Noeska Smit},
  journal = {Computers \& Graphics},
volume = {107},
pages = {208--219},
year = {2022},
ImageWebsite = {images/Sugathan_2022.png},
PDFWebsite = {https://www.sciencedirect.com/science/article/pii/S0097849322001479},
VideoWebsite={https://youtu.be/uwcqSf1W-dc},
doi = {10.1016/j.cag.2022.07.023}
}

@inproceedings{Sugathan_2021_VCBM,
booktitle = {Eurographics Workshop on Visual Computing for Biology and Medicine},
title = {{Interactive Multimodal Imaging Visualization for Multiple Sclerosis Lesion Analysis}},
author = {Sugathan, Sherin and Bartsch, Hauke and Riemer, Frank and Grüner, Renate and Lawonn, Kai and Smit, Noeska N.},
year = {2021},
publisher = {Eurographics Association},
ISSN = {2070-5786},
ISBN = {978-3-03868-140-3},
DOI = {10.2312/vcbm.20211346}
}

@inproceedings{Liu2008,
author = {Liu, Ligang and Zhang, Lei and Xu, Yin and Gotsman, Craig and Gortler, Steven J.},
title = {{A Local/Global Approach to Mesh Parameterization}},
year = {2008},
publisher = {Eurographics Association},
address = {Goslar, Germany},
booktitle = {Proceedings of the Symposium on Geometry Processing},
pages = {1495--1504},
numpages = {10},
location = {Copenhagen, Denmark},
doi ={10.1111/j.1467-8659.2008.01290.x}
}

@article{liu2015visual,
  title={{Visual Perception of Procedural Textures: Identifying Perceptual Dimensions and Predicting Generation Models}},
  author={Liu, Jun and Dong, Junyu and Cai, Xiaoxu and Qi, Lin and Chantler, Mike},
  journal={PLOS ONE},
  volume={10},
  number={6},
  pages={e0130335},
  year={2015},
  publisher={Public Library of Science San Francisco, CA USA},
  doi = {10.1371/journal.pone.0130335}
}

@INPROCEEDINGS{Interrante1996,
  author={Interrante, V. and Fuchs, H. and Pizer, S.},
  booktitle={Proceedings of Seventh Annual IEEE Visualization '96}, 
  title={{Illustrating Transparent Surfaces with Curvature-Directed Strokes}}, 
  year={1996},
  volume={},
  number={},
  pages={211--218},
  doi={10.1109/VISUAL.1996.568110}}

@ARTICLE{Isenberg2003,
  author={Isenberg, T. and Freudenberg, B. and Halper, N. and Schlechtweg, S. and Strothotte, T.},
  journal={IEEE Computer Graphics and Applications}, 
  title={{A Developer's Guide to Silhouette Algorithms for Polygonal Models}}, 
  year={2003},
  volume={23},
  number={4},
  pages={28--37},
  doi={10.1109/MCG.2003.1210862}}

@InBook{Lawonn_2015_Feature,
Title = {{Feature Lines for Illustrating Medical Surface Models: Mathematical Background and Survey}},
Author = {Kai Lawonn and Bernhard Preim},
Chapter = {Visualization in Medicine in Life Sciences III},
Pages = {93--132},
Publisher = {Springer},
Year = {2016},
doi = {10.1007/978-3-319-24523-2_5}
}

@inproceedings{HertzmannZorin2000,
author = {Hertzmann, Aaron and Zorin, Denis},
title = {{Illustrating Smooth Surfaces}},
year = {2000},
isbn = {1581132085},
publisher = {ACM Press/Addison-Wesley Publishing Co.},
address = {USA},
url = {https://doi.org/10.1145/344779.345074},
doi = {10.1145/344779.345074},
booktitle = {Proceedings of the 27th Annual Conference on Computer Graphics and Interactive Techniques},
pages = {517--526},
numpages = {10},
}

@article{Lawonn2013,
author = {Lawonn, K. and Moench, T. and Preim, B.},
title = {{Streamlines for Illustrative Real-Time Rendering}},
journal = {Computer Graphics Forum},
volume = {32},
pages = {321--330},
doi = {https://doi.org/10.1111/cgf.12119},
url = {https://onlinelibrary.wiley.com/doi/abs/10.1111/cgf.12119},
eprint = {https://onlinelibrary.wiley.com/doi/pdf/10.1111/cgf.12119},
year = {2013}
}

@article{Gerl2013,
title = {{Interactive Example-Based Hatching}},
journal = {Computers \& Graphics},
volume = {37},
number = {1},
pages = {65--80},
year = {2013},
issn = {0097-8493},
doi = {https://doi.org/10.1016/j.cag.2012.11.003},
url = {https://www.sciencedirect.com/science/article/pii/S0097849312001744},
author = {Moritz Gerl and Tobias Isenberg},
}

@article{Lawonn2014,
author = {Lawonn, Kai and Krone, Michael and Ertl, Thomas and Preim, Bernhard},
title = {{Line Integral Convolution for Real-Time Illustration of Molecular Surface Shape and Salient Regions}},
journal = {Computer Graphics Forum},
volume = {33},
number = {3},
pages = {181--190},
doi = {https://doi.org/10.1111/cgf.12374},
url = {https://onlinelibrary.wiley.com/doi/abs/10.1111/cgf.12374},
eprint = {https://onlinelibrary.wiley.com/doi/pdf/10.1111/cgf.12374},
year = {2014}
}

@article{Kruger2007,
  title={{Efficient Stipple Rendering}},
  author={Kr{\"u}ger, Jens and Westermann, R{\"u}diger},
  journal={Proceedings of IADIS Computer Graphics and Visualization},
  pages={19--26},
  year={2007},
  publisher={ADIS Lisbon}
}

@article{Martin2017,
title = {{A Survey of Digital Stippling}},
journal = {Computers \& Graphics},
volume = {67},
pages = {24--44},
year = {2017},
issn = {0097-8493},
doi = {https://doi.org/10.1016/j.cag.2017.05.001},
url = {https://www.sciencedirect.com/science/article/pii/S0097849317300432},
author = {Domingo Martín and Germán Arroyo and Alejandro Rodríguez and Tobias Isenberg},
}

@InProceedings{Lawonn_2014_BVM,
Title = {{Illustrative Visualization of Endoscopic Views}},
Author = {Kai Lawonn and Patrick Saalfeld and Bernhard Preim},
Booktitle = {Bildverarbeitung für die Medizin (BVM)},
doi = {10.1007/978-3-642-54111-7_52},
Year = {2014},
Pages = {276--281},
publisher= {Springer},
address={Berlin, Heidelberg, Germany}
}

@InProceedings{Lawonn_2015_MICCAI,
Title = {{Illustrative Visualization of Vascular Models for Static 2D Representations}},
Author = {Kai Lawonn and Maria Luz and Bernhard Preim and Christian Hansen},
Booktitle = {International Conference on Medical Image Computing and Computer Assisted Intervention (MICCAI)},
Year = {2015},
Address = {Munich, Germany},
Pages = {399--406},
doi = {10.1007/978-3-319-24571-3_48}
}

@ARTICLE{Taylor2002,
  author={Taylor, R.},
  journal={IEEE Computer Graphics and Applications}, 
  title={{Visualizing Multiple Fields on the Same Surface}}, 
  year={2002},
  volume={22},
  number={3},
  pages={6--10},
  doi={10.1109/MCG.2002.999781}}

@article{Martin2019,
  doi = {10.1016/j.cag.2019.02.001},
  url = {https://doi.org/10.1016/j.cag.2019.02.001},
  year = {2019},
  publisher = {Elsevier {BV}},
  volume = {80},
  pages = {1--16},
  author = {Domingo Mart{\'{\i}}n and Germ{\'{a}}n Arroyo and Vicente del Sol and Celia Romo and Tobias Isenberg},
  title = {{Analysis of Drawing Characteristics For Reproducing Traditional Hand-Made Stippling}},
journal = {Computers \& Graphics},
}

@ARTICLE{Sajadi2013,
  author={Sajadi, Behzad and Majumder, Aditi and Oliveira, Manuel M. and Schneider, Rosália G. and Raskar, Ramesh},
  journal={IEEE Transactions on Visualization and Computer Graphics}, 
  title={{Using Patterns to Encode Color Information for Dichromats}}, 
  year={2013},
  volume={19},
  number={1},
  pages={118--129},
  doi={10.1109/TVCG.2012.93}}

@ARTICLE{Lloyd1982,
  author={Lloyd, S.},
  journal={IEEE Transactions on Information Theory}, 
  title={{Least Squares Quantization in PCM}}, 
  year={1982},
  volume={28},
  number={2},
  pages={129--137},
  doi={10.1109/TIT.1982.1056489}}

@book{Cunningham2011,
    doi = {10.1201/b11308},
author = {Cunningham, Douglas and Wallraven, Christian},
title = {{Experimental Design: From User Studies to Psychophysics}},
year = {2011},
isbn = {1568814682},
publisher = {A. K. Peters, Ltd.},
address = {USA},
edition = {1st}
}

@inproceedings{cunningham2007perceptual,
  title={{Perceptual Reparameterization of Material Properties.}},
  author={Cunningham, Douglas W and Wallraven, Christian and Fleming, Roland W and Stra{\ss}er, Wolfgang},
  pages={89--96},
  year={2007},
  booktitle = {Computational Aesthetics in Graphics, Visualization, and Imaging},
publisher = {Eurographics Association},
address = {Banff, Alberta, Canada},
DOI = {10.2312/COMPAESTH/COMPAESTH07/089-096}
}

@article{lettvin1976,
  title={{On Seeing Sidelong}},
  author={Lettvin, Jerome Y and others},
  journal={The Sciences},
  volume={16},
  number={4},
  pages={10--20},
  year={1976},
  doi = {10.1002/j.2326-1951.1976.tb01231.x}
}

@article{caelli1978perceptual1,
  doi = {10.1007/bf00337138},
  url = {https://doi.org/10.1007/bf00337138},
  year = {1978},
  publisher = {Springer},
  volume = {28},
  number = {3},
  pages = {167--175},
  author = {T. Caelli and B. Julesz},
  title = {{On Perceptual Analyzers Underlying Visual Texture Discrimination: Part I}},
  journal = {Biological Cybernetics}
}

@article{julesz1981,
  title={{Textons, the Elements of Texture Perception, and their Interactions}},
  author={Julesz, Bela},
  journal={Nature},
  volume={290},
  number={5802},
  pages={91--97},
  year={1981},
doi = {10.1038/290091a0},
  publisher={Nature Publishing Group UK London}
}

@article{julesz1973inability,
  title={{Inability of Humans to Discriminate between Visual Textures that Agree in Second-Order Statistics—Revisited}},
  author={Julesz, Bela and Gilbert, Edgar N and Shepp, Larry A and Frisch, Harry L},
  journal={Perception},
  volume={2},
  number={4},
  pages={391--405},
  year={1973},
  doi={10.1068/p020391},
  publisher={SAGE Publications Sage UK: London, England}
}

@article{liu2010construction,
  title={{Construction of Iso-Contours, Bisectors, and Voronoi Diagrams on Triangulated Surfaces}},
  author={Liu, Yong-Jin and Chen, Zhanqing and Tang, Kai},
  journal={IEEE Transactions on Pattern Analysis and Machine Intelligence},
  volume={33},
  number={8},
  pages={1502--1517},
  year={2010},
  publisher={IEEE},
  doi = {10.1109/TPAMI.2010.221}
}

@article{Kabsch1976,
  doi = {10.1107/s0567739476001873},
  url = {https://doi.org/10.1107/s0567739476001873},
  year = {1976},
  publisher = {International Union of Crystallography ({IUCr})},
  volume = {32},
  number = {5},
  pages = {922--923},
  author = {W. Kabsch},
  title = {{A Solution for the Best Rotation to Relate two Sets of Vectors}},
  journal = {Acta Crystallographica Section A}
}

@article{Wills2009,
  doi = {10.1145/1559755.1559760},
  url = {https://doi.org/10.1145/1559755.1559760},
  year = {2009},
  publisher = {Association for Computing Machinery ({ACM})},
  volume = {28},
  number = {4},
  pages = {1--15},
  author = {Josh Wills and Sameer Agarwal and David Kriegman and Serge Belongie},
  title = {{Toward a Perceptual Space for Gloss}},
  journal = {{ACM} Transactions on Graphics}
}

@misc{Timofeev2010,
  doi = {10.2210/pdb3i40/pdb},
  url = {https://doi.org/10.2210/pdb3i40/pdb},
  year = {2010},
  publisher = {Worldwide Protein Data Bank},
  author = {V.I. Timofeev and V.V. Bezuglov and K.A. Miroshnikov and R.N. Chuprov-Netochin and I.P. Kuranova},
  title = {{Human Insulin}}
}

@article{Timofeev2010a,
  doi = {10.1107/s1744309110000461},
  url = {https://doi.org/10.1107/s1744309110000461},
  year = {2010},
  publisher = {International Union of Crystallography ({IUCr})},
  volume = {66},
  number = {3},
  pages = {259--263},
  author = {V. I. Timofeev and R. N. Chuprov-Netochin and V. R. Samigina and V. V. Bezuglov and K. A. Miroshnikov and I. P. Kuranova},
  title = {{X-ray Investigation of Gene-Engineered Human Insulin Crystallized from a Solution Containing Polysialic Acid}},
  journal = {Acta Crystallographica Section F Structural Biology and Crystallization Communications}
}

@misc{uisBrodatzTextures,
	author = {Trygve Randen},
	title = {{B}rodatz {T}extures},
	howpublished = {\url{https://www.ux.uis.no/~tranden/brodatz.html}},
	year = {},
	note = {[Accessed 01-Apr-2023]},
}

@ARTICLE{Acevedo2006,
  author={Acevedo, Daniel and Laidlaw, David},
  journal={IEEE Transactions on Visualization and Computer Graphics}, 
  title={{Subjective Quantification of Perceptual Interactions among some 2D Scientific Visualization Methods}}, 
  year={2006},
  volume={12},
  number={5},
  pages={1133--1140},
  doi={10.1109/TVCG.2006.180}}

@article{Meuschke_2017_TVCG,
Title = {{Combined Visualization of Vessel Deformation and Hemodynamics in Cerebral Aneurysms}},
Author = {Monique Meuschke and Samuel Voß and Oliver Beuing and Bernhard Preim and Kai Lawonn},
Year = {2017},
Pages = {761--770},
Volume = {23(1)},
Journal = {IEEE Transactions on Visualization and Computer Graphics},
doi={https://doi.org/10.1109/TVCG.2016.2598795},
}



\end{document}